\documentclass[useAMS,usenatbib]{mn2e}
\usepackage{amsmath}
\usepackage{amsfonts}
\usepackage{amssymb}
\usepackage{natbib}
\usepackage{graphicx}
\usepackage{multirow}
\usepackage{textcomp}
\usepackage[normalem]{ulem}
\useunder{\uline}{\ul}{}
\usepackage{upgreek}
\usepackage{hyperref}
\usepackage{subfig}
\hypersetup{
        colorlinks=true,
        linkcolor=red,
        citecolor=blue,
        filecolor=magenta,
        urlcolor=cyan
        }

\newcommand\PHI{\texttt{PHI}}

 \include{definitions}

\makeatletter

\newcommand{\Rmnum}[1]{\expandafter\@slowromancap\romannumeral #1@}
\makeatother

\title[Bayesian bulge-disc decomposition]{Bayesian bulge-disc decomposition of galaxy images}
\author[J. J. Argyle et al.]{J. J. Argyle$^{1}$\thanks{E-mail:
ja66@st-andrews.ac.uk}, J. M\'{e}ndez Abreu$^{1,2,3}$, V. Wild$^{1}$ and D. J. Mortlock$^{4,5,6}$ \\
$^1$  School of Physics \& Astronomy, University of St Andrews, North Haugh, St Andrews, KY16 9SS, Scotland \\
$^2$ Instituto de Astrofísica de Canarias, Calle Vía Láctea s$/$n, E-38205 La Laguna, Tenerife, Spain \\
$^3$ Departamento de Astrofsica, Universidad de La Laguna, E-38206, La Laguna, Spain \\
$^4$ Astrophysics Group, Imperial College London, Blackett Laboratory, Prince Consort Road, London SW7 2AZ, UK \\
$^5$ Statistics Section, Department of Mathematics, Imperial College London, London SW7 2AZ, UK \\
$^6$ Department of Astronomy, Stockholm University, Albanova, SE-10691 Stockholm, Sweden}
\begin{document}

\pagerange{\pageref{firstpage}--XX} \pubyear{2017}

\maketitle

\label{firstpage}

\begin{abstract}
We introduce \PHI, a fully Bayesian Markov-chain Monte Carlo algorithm designed for the structural decomposition of galaxy images. \PHI\ uses a triple layer approach to effectively and efficiently explore the complex parameter space. Combining this with the use of priors to prevent nonphysical models, \PHI\ offers a number of significant advantages for estimating surface brightness profile parameters over traditional optimisation algorithms. We apply \PHI\ to a sample of synthetic galaxies with SDSS-like image properties to investigate the effect of galaxy properties on our ability to recover unbiased and well constrained structural parameters. In two-component bulge+disc galaxies we find that the bulge structural parameters are recovered less well than those of the disc, particularly when the bulge contributes a lower fraction to the luminosity, or is barely resolved with respect to the pixel scale or PSF. There are few systematic biases, apart from for bulge+disc galaxies with large bulge S\'{e}rsic parameter, $n$. On application to SDSS images, we find good agreement with other codes, when run on the same images with the same masks, weights, and PSF. Again, we find that bulge parameters are the most difficult to constrain robustly. Finally, we explore the use of a Bayesian Information Criterion (BIC) method for deciding whether a galaxy has one- or two-components.
\end{abstract}

\begin{keywords}
methods:data analysis -- methods:statistics --techniques:image processing -- techniques:photometric -- galaxies:photometry -- galaxies:structure
\end{keywords}

\section{Introduction}
Galaxy morphologies are complex, arising from many different processes that dictate the formation and evolution of the galaxy as a whole. Accurately characterising galaxy structure, \textit{i.e.} bulges, discs, bars and further complex components, is crucial for furthering our understanding. 

The human brain is extremely adept at pattern recognition and the classification of galaxy images began with the `Hubble Tunning Fork' \citep{Hub36,San61}. Today visual classification is still widely used and has been recently revitalised by the Galaxy Zoo project enabling amateur galaxy classifiers from across the globe to process vast amounts of galaxy structural information \citep{Lin08,Lin11,Wil13,Sim17}. Furthermore, the introduction of new techniques to mimic how the human brain captures the full, complex distribution of light has further advanced the usefulness of visual classification \citep{Hue08,Hue15,Die15}. 

An alternative and complementary approach is through the quantitative description of galaxy structures, using either parametric or non-parametric methods. Examples of non-parametric classifiers are concentration, clumpiness and asymmetry \citep{Con03,Lot04,Paw16}. Parametric methods include  S\'{e}rsic profile fitting \citep{Ser68} and multi-Gaussian expansion \citep{Ben91,Fas98,Ems98,Ode02,Cap02}. S\'{e}rsic profile fitting has become increasingly popular in recent decades due to its ability to reproduce the basic structures of many nearby and distant galaxies with typically only one or two (bulge and disc) components. Initially ellipticity-averaged 1D surface brightness profiles were used to fit the photometric components. However, this was shown to lead to systematic errors, as it does not account for the intrinsic shapes or position angle of the bulges \citep{Kor77,Bor81}, and most modern studies now fit the 2D images pixel-by-pixel.   

There are a large number of 2D fitting algorithms: GIM2D \citep{Sim98}, GALFIT \citep{Pen02, Pen10}, BUDDA \citep{Des04}, GASP2D \citep{Men08,Men17}, IMFIT \citep{Erw15}, GALPHAT \citep{Yoo11} and PROFIT \citep{Rob17}. Some of these codes (e.g., GALFIT and GASP2D) use minimisation algorithms to efficiently search for the best solution using the gradient of the model with respect to the parameters. Although these frequentist algorithms have commonly been employed to solve multi-component, non-linear fits, they suffer from some important drawbacks when faced with a problem as complex as photometric decomposition. \citet{Lan16} listed five commonly occurring factors which lead to failure of the \textit{Levenberg-Marquardt} (LM) fitting algorithm: i) Local minima trapping; ii) unrealistic solutions; iii) reversal of components \citep{All06}; iv) indecisiveness as to which model to use; and v) bad representation of final errors. To avoid some of these issues, they advocated the use of a grid of starting values combined with a convergence test to obtain robust parameter values. Typically 20\% to 30\% of automatic fits are deemed physically unrealistic; previous studies have often employed logical filters to weed these out  \citep[\textit{e.g. }][]{All06,Sim11,Mee15,Men17}. 

To circumvent these difficulties,  a more modern approach is to embed the galaxy morphology analysis into the broader context of inference and hypothesis testing with the use of Bayesian inference. The above problems can then be solved in turn: the exploration of parameter space can overcome runs that become trapped in local minima; initial priors can prevent unrealistic solutions and the reversal of components; model comparison tests can help determine the most probable morphology; and the posterior distribution gives a proper description of the parameter uncertainties. GALPHAT, PROFIT and IMFIT (version 1.4) offer a Markov Chain Monte Carlo (MCMC) approach to help overcome these problems.     

This paper introduces a new adaptive Bayesian MCMC algorithm, which has been purpose designed to obtain robust galaxy morphologies from galaxy images. We demonstrate its use on 2-component bulge-disc decomposition of both synthetic and real galaxy images. The aim is to provide a flexible, open source code in which it is simple for users to define their own models, PSFs, priors and likelihoods. The code is available for download in IDL (Interactive Data Language, \url{https://github.com/SEDMORPH/PHI/}). A Python version is also under developement.

In Section \ref{sec:method}, we describe the basic formalism of the inference methodology, including an overview of Bayesian statistics and details of the priors used. In Section \ref{sec:synthetic} we apply the method to an ensemble of synthetic galaxy images and discuss the interpretation of the outputs. In Section \ref{sec:data} the algorithm is applied to real galaxies within the Sloan Digital Sky Survey. In Section \ref{sec:BIC} we investigate a method to formally compare the one- and two-component model fits.  Finally, in Section \ref{sec:summary} we discuss and summarise the results of this paper.

\section{Inference Methodology}\label{sec:method}
\begin{figure}
\centering 
\includegraphics[width=0.65\textwidth, page=1, trim=0 0 0 0, clip]{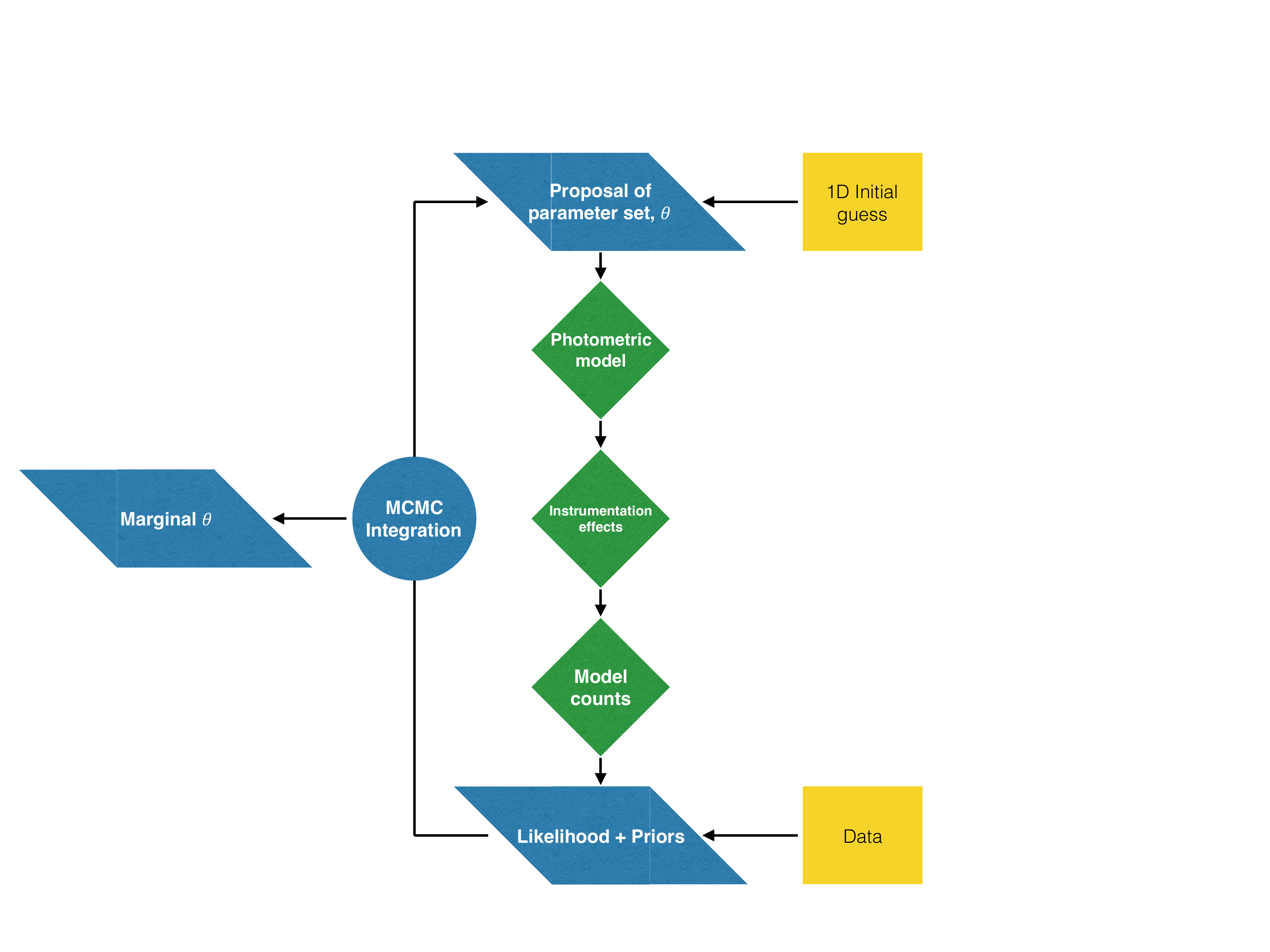}
\caption{The flow chart of the code \PHI\ (2D PHotometric decompositions using Bayesian Inference).}
\label{fig:flow}
\end{figure}

In this section we describe the main attributes of the code \PHI\ (2D PHotometric decompositions using Bayesian Inference). The key steps of our inference method to perform 2D PHotometric decompositions, illustrated in Fig.~\ref{fig:flow}, are:
\begin{enumerate}
\item \PHI\ reads in the flux and error maps of an image in \texttt{FITS} format. A point-spread-function (PSF) must either be provided as an image or specified in functional form. 
\item The user provides functions describing the components they wish to fit (Section \ref{sec:model}) and defines the priors. Initial guesses for the model parameters may be provided, but are not essential for the algorithm to function correctly.  
\item \PHI\ simulates the galaxy image with the chosen model and initial model parameters, and uses a fast Fourier transform to convolve the simulated image with the PSF.
\item The likelihood and posterior probability are calculated for the model and data.  
\item The MCMC engine commences by iteratively updating the model parameters and repeating steps 3-4 until the full posterior distribution has been mapped, or the user defined maximum iteration number is reached.  
\end{enumerate}

\subsection{2D-Photometric model functions}\label{sec:model}
\PHI\ has been designed to easily fit any parametric model to a galaxy image. However, for simplicity, in this paper we focus on the two main observed components of galaxies: spheroids and discs. In general, discs are well described by an exponential profile \citep{Vau56, Free70}, with the intensity $I$ changing with radius $R$ as, 
\begin{equation}
I(R) = I_{0} \, \exp \left(-\frac{R}{h} \right),
\end{equation}  
where $I_{0}$ is the central intensity and $h$ is the disc scale length. The spheroid component of galaxies can be modelled by S\'{e}rsic's (\citeyear{Ser68}) generalization of de Vaucouleurs' (\citeyear{Vau48,Vau56}) $R^{1/4}$ function to give the $R^{1/n}$ surface density profile, 
\begin{equation}
I(R) = I_{e} \, \exp \left\{ -b_{n}  \left[ \left(\frac{R}{R_{e}} \right )^{1/n} - 1 \right] \right\},
\end{equation}
where $I_{e}$ is the intensity at the effective radius $R_{e}$ that encloses half of the total light from the model and $n$, the  S\'{e}rsic index, describes the concentration of the light profile. The final parameter, $b_{n}$, is specified by $n$. When $n = 1$ the model follows an exponential surface-intensity profile and $n = 4$ reproduces the  de Vaucouleurs' model; thus the S\'{e}rsic profile can describe the two main observed components of galaxies. A detailed review of the S\'{e}rsic profile and associated quantities is given by \citet{Gra05}. 

We characterise the intensity profile of a galaxy by concentric elliptical isophotes with position angle $\theta_{\textrm{PA}}$ in degrees counter-clockwise from the vertical axis of the image, and ellipticity $\epsilon = 1 - q$, where $q= b/a$ is the ratio between the semi-minor and semi-major axis of the ellipse. The projected radius is given by,  
\begin{equation}
 r = x_p^2 + \frac{y_p^2}{q^2}
\end{equation}
where $x_p$ and $y_p$ are coordinates in the reference frame centred on the image centre $(x_0,y_0)$ and rotated to the position angle relative to the image $x$-axis ($\textrm{PA}=\theta_{\textrm{PA}} + 90\deg$),

\begin{equation}
\begin{aligned}
x_{p} &= (x - x_{0}) \, \cos \, (\textrm{PA}) \, + (y-y_{0}) \, \sin \, (\textrm{PA}) \\
y_{p} &= -(x - x_{0}) \, \sin \, (\textrm{PA}) \, + (y-y_{0}) \, \cos \, (\textrm{PA}).
\end{aligned}
\end{equation} 

Many previous studies have shown that a careful analysis of the PSF is needed to perform robust photometric decompositions. \citet{Men08} found that errors of $\sim$2\% in the PSF full width half maximum (FWHM) led to errors of up to 10\% in the $R_e$ and $n$ of the bulge. \citet{Gad09} found that to reliably retrieve the structural properties of bulges, the effective radius must be larger than $\sim$80\% of the PSF FWHM. It is therefore crucial that an accurate model for the PSF is provided to \PHI, and that \PHI\ then treats the PSF correctly. Given the large number of models that must be built for comparison with the data during the running of \PHI, we explored a variety of methods for convolving the model image with the PSF. We found that the Fast Fourier Transform (FFT) returns the required accuracy for the decomposition of images. 

\subsection{Bayesian framework}
Bayesian methods combine prior knowledge about a model with data to obtain a probabilistic description of the model. This is described by Bayes Theorem, 
\begin{equation}\label{eqn:bayes}
p(\theta|D) = \frac{L(D|\theta) p(\theta)}{\int L(D|\theta)p(\theta)d\theta},		
\end{equation}
where the model is characterised by the parameter vector $\theta$, $p(\theta|D)$ is the posterior probability of a set of parameters $\theta$ given the data $D$; $L(D|\theta)$ is the likelihood function or the probability of the data given $\theta$; and $p(\theta)$ is the prior probability of the parameter vector $\theta$.  The denominator ensures that the probability is unity when summed over all possible models.  

\subsubsection{Likelihood}
For a large number of photons detected in each independent CCD pixel, the measurement errors can be considered to be Gaussian with mean zero and no covariance. 
The likelihood for a given pixel $i$ is then given by,
\begin{equation}
p(d_{i}|\theta) = \frac{1}{(2 \pi \sigma_i^2)^{1/2}} \exp \left\{  -\frac{1}{2} \left[ \frac{d_i - f(x_i; \theta)}{\sigma_i} \right ]^2 \right \}
\end{equation}
where $f(x;\theta)$ is the model function consisting of known quantities $x$ (\textit{i.e}, constants, control variables, etc.) and the unknown parameters $\theta$.
Since each pixel is considered to be an independent measurement, the combined $-2 \ln$ likelihood of all the $N$ pixels is,
\begin{equation}
 -2\ln p(D|\theta) = \chi^2 + \sum_{i=1}^N \ln \sigma_{i}^{2} + N \ln 2 \pi
\end{equation}  
where
\begin{equation}
\chi^2 = \sum_{i=1}^N \frac{[d_i - f(x_i; \theta)]^2 }{\sigma_i^2}.
\end{equation}

\subsubsection{Priors}\label{sec:priors}

\begin{table*}
\centering
\caption{Details of the prior distributions set for each of the model parameters used in this paper. $U[a,b]$ specifies a uniform distribution between lower and upper limits $a$ and $b$ and $\delta(a)$ specifies a Kronecker delta function with probability of 1 at $a$ and 0 otherwise. $S_{\textit{image}}$ is the size of the image.}
\label{tab:priors}
\begin{tabular}{llll}
\hline
\textbf{Individual parameters}   &                &                             &                                               \\
  Parameter                         & Symbol        & Prior                                  & Range                                                       \\ \hline
  Effective intensity            & $I_{e}$          & Uniform in $\log (I_{e})$    & $\log (I_{e} / counts) \in U[0.01,10]$        \\
  Effective radius                & $R_{e}$         & Uniform in $\log (R_{e})$   & $\log (R_{e} / pixels) \in U[0.01,10]$        \\
  S\'{e}rsic index                 & $n$             & Uniform in $n$                  & $n \in U[0.4, 8]$                                       \\
  Central intensity              & $I_{0}$          & Uniform in $\log (I_{0})$    & $\log (I_{0} / counts) \in U[0.01,10]$        \\
  Scale length                     & $h$             & Uniform in $\log (h)$         & $\log (h / pixels) \in U[0.01,10]$            \\
  Axial ratio                        & $q$             & Uniform in $q$                  & $q \in U[0.2, 1]$                                       \\
  Position angle                  & PA               & Uniform in PA                    & PA/degrees $\in U[-360, 360]$                 \\
  Central coordinates         & $x_{0},y_{0}$ & Uniform in $x_{0} \& y_{0}$ & $x_{0} \& y_{0} \in U[0, S_{\textit{image}}]$     \\ \hline

\textbf{Combined parameters}     &                &                             &                                              \\ \hline
Effective radius / scale length  & $R_{e} / h$      & Uniform in $R_{e} / h$        & $R_{e} / h \in U[0.05, 1.678]$       \\
Bulge-to-total flux ratio                  & $B/T$             & Uniform in $B/T$               & $B/T \in U[0.01, 1]$                     \\
Bulge-to-disc flux ratio $(< R_{e})$ & $B/D(< R_{e})$ & Uniform in $B/D(< R_{e})$   & $B/D(< R_{e}) \in U[1, -]$             \\
\# of crossing points                & $N_{x}$           & $\delta(N_x)=1$ for $N_x=1$       &                                                    \\ \hline
\\
\end{tabular}
\end{table*}

\begin{figure*}
\centering 
\includegraphics[width=1\textwidth, page=5, trim=0 140 100 100, clip ]{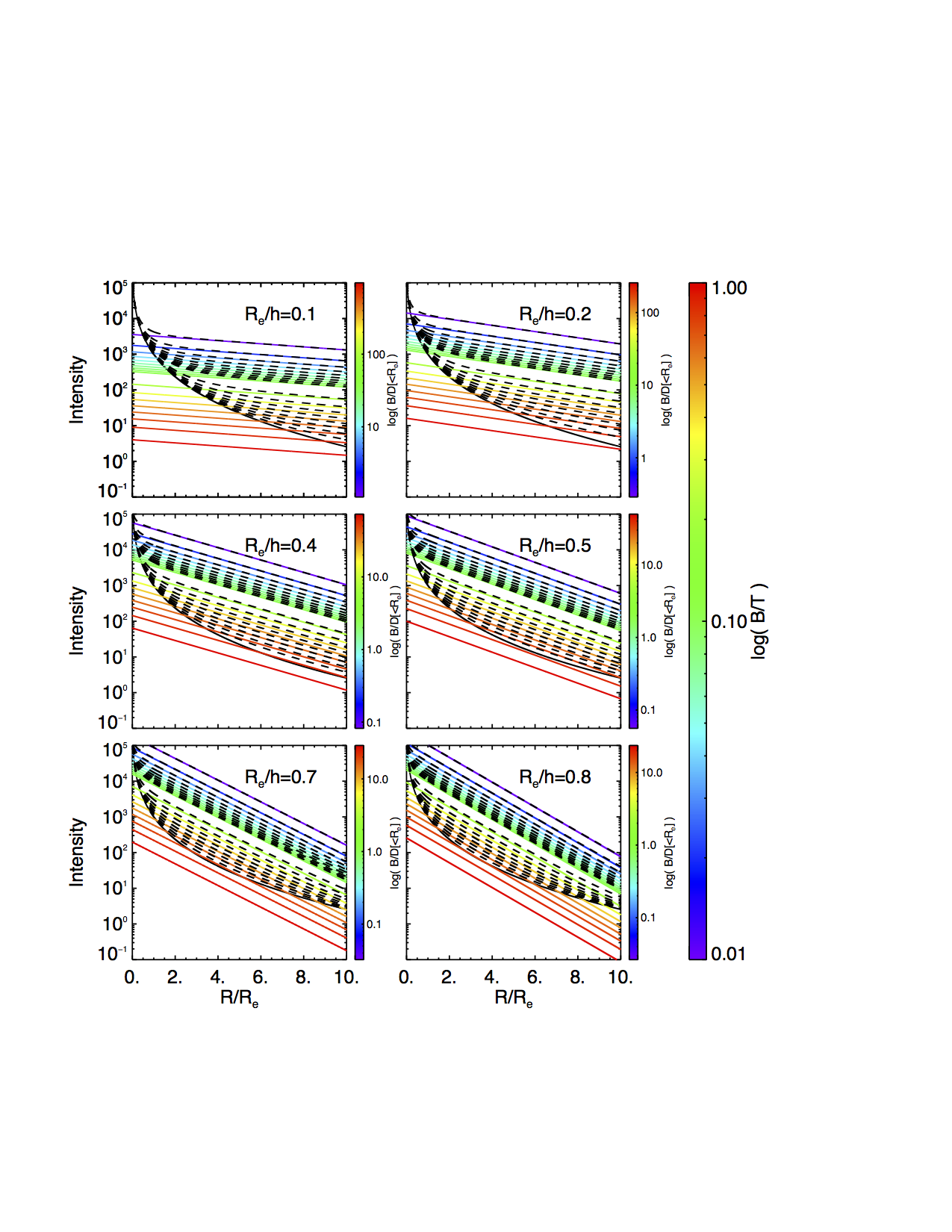}
\caption{The intensity profiles for two-component models, where the radius has been normalised by the effective radius of the bulge. The solid black line shows the bulge profile in the case of a S\'{e}rsic model with $n=4$. The coloured lines show the exponential disc component, with different bulge-to-total flux ratios ($B/T$) as given in the large colour bar, or bulge-to-disc flux ratios ($B/D$) within one effective radius as shown by the individual colour bars. The black dashed lines show the combined (total) intensity profiles. Each panel shows models with different ratios of bulge effective radius to disc scale length ($R_{e} / h$) as given in the legend.}
\label{fig:priors}
\end{figure*}

The prior distribution expresses our knowledge and prejudices about certain values of the parameters, or relationships between them. Where we know some values to be more probable than others, a carefully selected prior distribution can encode this knowledge. However, biases may arise if the prior distribution is \textit{informative} \textit{e.g.}, a Gaussian probability distribution function (PDF) with a narrow width. For the purpose of galaxy image analysis, where the parameter space is known to have many local minima, it is only advisable to use an informative prior where there is a clear justification. \PHI\ will accept any valid function for the prior distributions defined by the user. In this paper we demonstrate the use of \PHI\ with maximally \textit{uninformative} priors, i.e., a uniform distribution between certain parameter limits. 

The top section of Table \ref{tab:priors} lists the prior functions used in this paper for each parameter of the model in Section \ref{sec:model}. In the case of photometric decompositions of galaxy images there are some clearly physically motivated limits on the parameters. For example, negative or very large values for the radius are unphysical. We allow values for the S\'{e}rsic index, $n$, to take any value within the range $0.5 < n < 8$. Larger values produce unphysical concentrations of light in the centre. 

One advantage of the Bayesian inference framework for the photometric decomposition of {\it two-component} galaxy images is that the accidental fitting of a one-component model, or the reversal of bulge and disc components, can be explicitly avoided through the use of combined priors. To illustrate the use of combined priors, we restrict our models to fit a bulge defined to be an excess of light over the inner extrapolation of an exponential disc. We note that this particular model may not be appropriate for all science goals, for example in the study of embedded discs.  To do this we must constrain our two-component galaxies to be a combination of an {\it inner} S\'{e}rsic profile (the bulge) and {\it outer} exponential profile (the disc). Unfortunately, this is less straightforward than it sounds: for a bulge profile with S\'{e}rsic parameter $n > 1$ and an exponential disc profile, at some (large) radius the inner component will again dominate over the outer. To explore where the reversal of components could occur in practice Fig.~\ref{fig:priors} shows the effect on the intensity profiles of varying each of the model parameters presented in Section \ref{sec:model}, for a bulge with $n=4$. The bulge component dominates at large radii when the ratio of bulge effective radius to disc scale length ($R_{e} / h$) is large and the bulge-to-total flux ratio ($B/T$) is large. In this example, we employ combined priors to enforce our prejudice about the relative strength and positioning of the two components in three ways: i) preventing the disc effective radius ($R_{\textrm{disc}} = 1.678h$) becoming smaller than the bulge $R_e$; ii) ensuring the bulge-to-disc flux ratio ($B/D$) within one effective bulge radius is always larger than unity; iii) ensuring that the number of crossing points in the bulge and disc 1D light profiles ($N_x$) is unity. This third prior ensures that two components are fit, yet prevents the light profile of the bulge becoming dominant in the outer edges. It is implemented via a Newton-Raphson algorithm, which is run until the total model intensity falls below the mean of the sky background; after this a second crossing point can occur (and is inevitable for galaxies with $n > 1$). These combined priors are summarised in the second half of Table \ref{tab:priors}, where we again use maximally uninformative priors for simplicity. More complex priors may be included trivially by the user, depending on their science goals. 

\subsection{Model comparison}\label{sec:methodBIC}
The aim of model fitting is to construct probabilistic models that represent, or sufficiently approximate, the data. Once the simplest model has been fit, we can increase the complexity of the model by adding extra parameters. It is important to then test whether each additional parameter is justified on the grounds of a significantly improved fit, given the increased number of degrees of freedom. 

In this paper we investigate the use of the Bayesian Information Criteria \citep[\textrm{BIC}, ][]{Sch78}, which compares the maximum likelihood of each model $L(D|\theta_{\rm ML})$,  
\begin{equation}
\textrm{BIC} = -2 \ln(p(D|\theta_{\rm ML})) + m \ln(N)
\end{equation} 
where $N$ is the number of data points and $m$ is the number of free parameters in the model ($\theta_{\rm ML}$ is the corresponding maximum likelihood parameter vector). The difference between the BIC of two models ($A$ and $B$) is,
\begin{equation}
\Delta\textrm{BIC} = -2 \ln \left( \frac{p(D|\theta_{A})}{p(D|\theta_{B})} \right) + (m_{A} - m_{B}) \ln(N).
\end{equation}  
The actual calculation of the \textrm{BIC} requires the Bayes factor, which is not provided by the \PHI\ algorithm. We therefore approximate $\Delta\textrm{BIC}$ by fitting both models to the data and taking the posterior medians for each set of fitted parameters. Unfortunately, the appropriate demarcation to distinguish between the models is somewhat dependent on the problem in hand. We therefore carry out simulations in Section \ref{sec:BIC} to determine the appropriate values for our dataset.
 
\subsection{The MCMC engine}

\begin{figure}
\centering 
\includegraphics[width=0.65\textwidth, page=1, trim=55 150 50 150, clip]{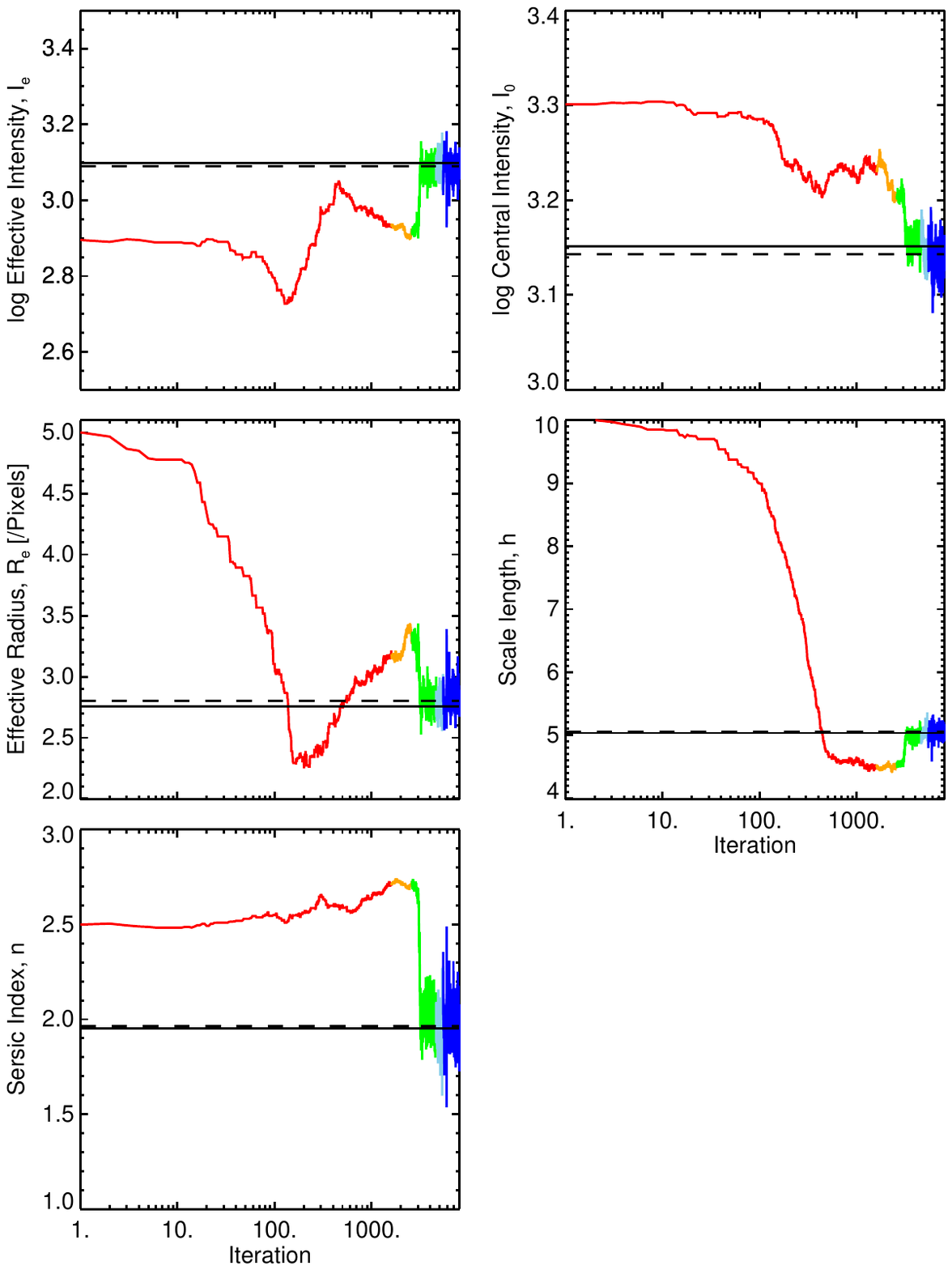}
\caption{The MCMC trace plots for  some of the profile parameters in a two-component model. The left column shows the bulge (S\'{e}rsic) profile parameters: the effective intensity ($I_{e}$), effective radius ($R_{e}$), and S\'{e}rsic index ($n$). The right column shows the disc (exponential) profile parameters: the central intensity ($I_{0}$) and scale length ($h$). The red, green and blue correspond to the three levels of the algorithm. The orange shows the transition between 1st and 2nd levels, and the light blue shows the burn-in part of the 3rd level that is discarded.}
\label{fig:chains}
\end{figure}

Exact Bayesian analysis is restricted by the need to perform integrations analytically. For a simple model with two or three parameters, one can obtain a good estimate of the posterior probability by exploring all possible parameter values on a grid. For a high-dimensional parameter space, such as the models used to describe the surface brightness distribution of galaxies, the characterisation of the posterior distribution becomes increasingly difficult, and the use of a grid is prohibitively time consuming. Sampling-based methods allow the exploration of highly multi-dimensional and complex parameter spaces. One example of these methods is Markov-chain Monte Carlo (MCMC): by generating repeated states by a first-order Markovian process, MCMC asymptotically converges to the posterior distribution. 

The purpose-built MCMC algorithm used in \texttt{PHI} consists of three levels that aim to achieve an efficient convergence and accurate estimation of the posterior distribution. Fig. \ref{fig:chains} shows a typical run of the entire algorithm with different colours depicting the three levels and transitions between them. In the following subsections we will address the intricacies of each level individually.

\subsubsection{Level one: Blocked Adaptive Metropolis}
\texttt{PHI} begins with a variation on the Adaptive-Metropolis-within-Gibbs algorithm introduced in \citet{Rob09}. The purpose of this level is to obtain an estimate of the scale of each parameter in the Markov chain. By knowing how each parameter scales \texttt{PHI} can efficiently sample from the parameter space, which overcomes problems with poorly chosen initial parameter values, as well as complex probability distributions with many local minima. Given a current value in the Markov chain, $X_{i}$, a new value or set of values $Y$ is proposed, where $i$ denotes the $i^{th}$ step in the Markov chain. The new values $Y$ are either accepted as a valid move so the next starting location is $X_{i+1} = Y$, or are rejected and $X_{i+1} = X_{i}$, according to the criteria,
\begin{equation}
\begin{cases} 
X_{i+1} = Y & \text{if  $ U < min [1,\pi (Y)/\pi (X_{i})  ] $} \\
X_{i+1} = X_{i} & \text{if $ U \geq min [1, \pi (Y)/\pi (X_{i})  ]$}
\end{cases}.
\end{equation}
where $U$ is a uniformly chosen random number $U \sim U(0,1)$ and $\pi (\centerdot )$ is the target distribution (i.e. the combination of likelihood and prior distributions in Eq.~\ref{eqn:bayes}). This process is repeated for every parameter sequentially. 

The proposed new parameters are drawn from a Gaussian function $Y \sim N(X_{i,j}, \sigma_{i,j}^{2})$, where $\sigma_{i,j}$ represents the size of the step the algorithm makes when choosing the proposed values at each iteration $i$ and for each parameter $j$. The correct value for $\sigma_{i,j}$ will provide a compromise between being able to jump from one region of parameter space to another quickly, and being able to explore in detail the target distribution. If $\sigma_{i,j}$ is too large we see a drop in the acceptance rate, as we are drawing from a region of parameter space with low probability. For a $\sigma_{i,j}$ which is too small we will accept values at almost every iteration. The optimal acceptance rate is 0.44 for a one-dimensional Markov chain and 0.23 for dimensions greater than one \citep{Rob01}. The Adaptive-Metropolis-within-Gibbs algorithm uses information from past iterations to adapt $\sigma_{i,j}$ until the desired acceptance rate is achieved. 

To accomplish diminishing adaptation we initially calculate the average acceptance rate for the past $n_{step}$ (default $n_{step} = 100$) of iterations and allow an update to the $\sigma_{i,j}$ by adding or subtracting 5$\%$ of $\sigma_{i,j}$ to adjust the acceptance rate closer to the optimal value \citep{Rob01}. Once the acceptance rate falls within 0.15 and 0.32, the average acceptance rate is calculated over the last $2n_{step}$ until every parameter (or blocked set of parameters) again has an average acceptance rate within 0.15 and 0.32. We then monitor the acceptance rate for a further $4n_{step}$ iterations, and adjust $\sigma_{i,j}$ until the acceptance rate falls between 0.15 and 0.32. At that point, adaptation is stopped and the final $\sigma_{i,j}$ values are saved. 

It is important that the final Markov chain closely matches the target distribution, so the chain is continued without any further adaptation of the $\sigma_{j}$ until the chain's gradient tends to zero. This is done by calculating the average parameter value after every 200 iterations, and once 5 averages are obtained a line is fit and the gradient of this line determined. When this gradient is close to zero the chain is converging to the target distribution and the algorithm can move onto Level two.

\subsubsection{Level two: Adaptive Metropolis}
The aim of this level is to obtain a similar covariance structure for the proposal distribution ($Y$) to that of the target distribution, which leads to greater success rates for the proposal distribution \citep{Haa01,Rob09}. $Y$ is drawn in a similar way to before: $Y \sim N(X_{i}, c \Sigma_{i})$, again where $X_{i}$ is the current state of the chain, and the same accept/reject Metropolis rule is used as in Level one. $\Sigma_{n}$ is the covariance matrix of all the previously generated values of the chain since the adaptation of level one finished, and $c$ is a constant that is included to yield an optimal acceptance rate: $c=2.382^{2} /d$ where $d$ is the number of parameter dimensions (see \citealt{Haa01} and \citealt{Rob09}). 

To establish if further adaptation will improve the chain the algorithm tests that the covariance structure of the target distribution has been correctly identified. This can be determined directly from the past iterations of the chain. Every $N_{L2}$ (user input) iterations the mean squared difference between each successive iteration $\langle (X_{i-1} - X_{i})^2 \rangle$ for each parameter is calculated, and after $5N_{L2}$ a linear model is fit. If the gradient of the mean squared differences appears to have an increasing or decreasing gradient then the algorithm continues to adapt; if the gradient is close to zero then adaptation stops and the algorithm moves to the final level.    

\begin{figure*}
\centering 
\includegraphics[width=1.\textwidth, page=1, trim=40 150 40 180, clip]{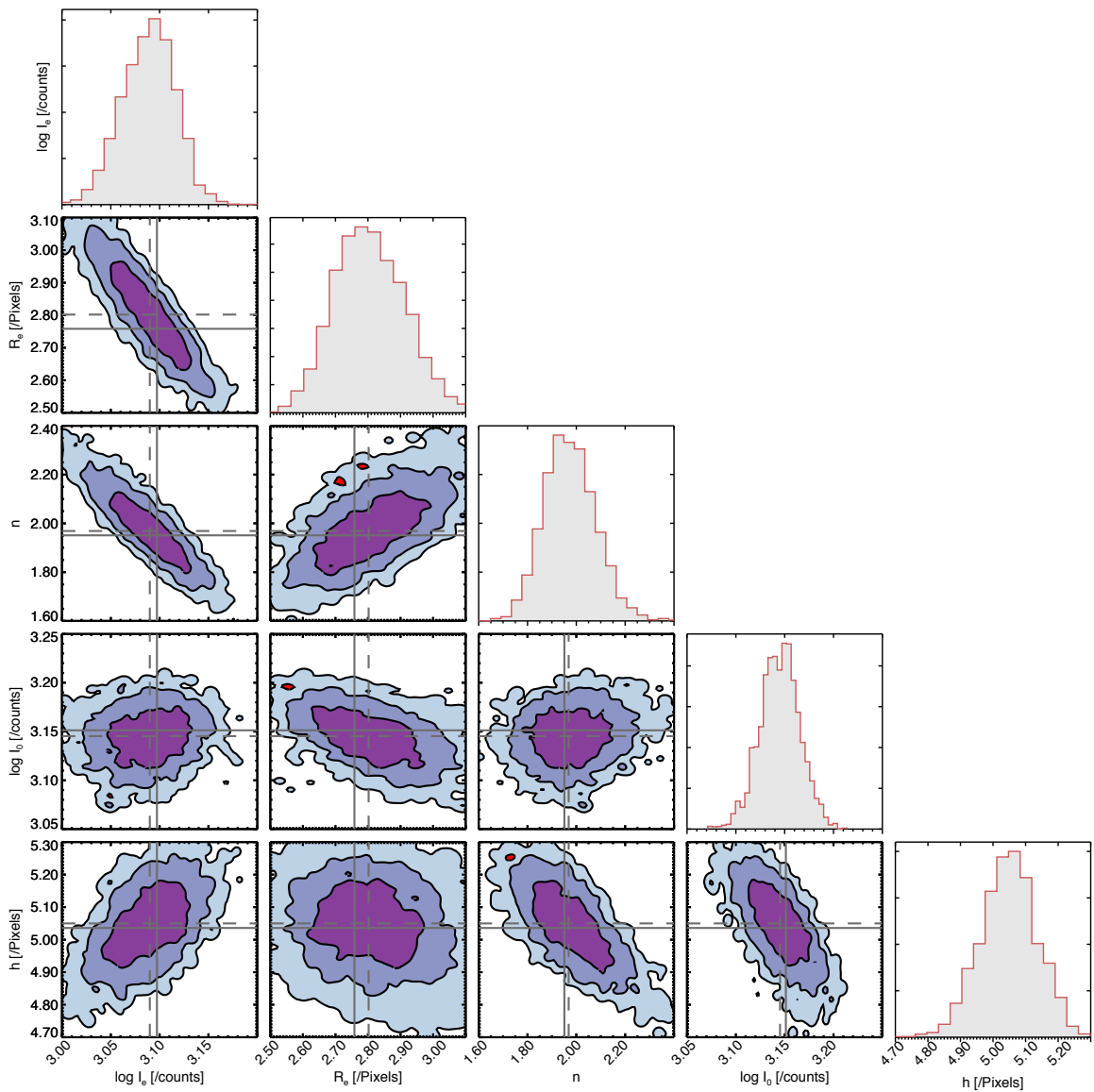}
\caption{Example marginal posterior distributions for a two-component synthetic SDSS $i$-band galaxy image. The marginal distribution for each of the structural parameters is shown on the diagonal. Joint marginal distributions for pairs of parameters are shown in the off-diagonal panels. The three color contours represent the 68, 95, and 99\% confidence levels. The solid grey line shows the true value of this synthetic galaxy and the dashed line indicates the median of the posterior distribution.}
\label{fig:postone}
\end{figure*}

\subsubsection{Level three \& chain convergence}
The final level of \texttt{PHI} involves a symmetric random walk Metropolis algorithm drawing the proposed values from $Y \sim N(X_{i}, c \Sigma_{L2})$, where $\Sigma_{L2}$ is the last covariance matrix calculated before adaptation stopped in Level 2 and $c$ is again the constant to help achieve the target acceptance rate.  In \texttt{PHI} the default method to test for convergence is to run multiple chains simultaneously, and then to use a Gelman-Rubin diagnostic \citep{Gel92}. Alternatively, a Geweke diagnostic \citep{Gew92} can be used to determine whether a single Markov chain has converged. Once the Markov chains have converged, the chains are combined to form the final sample distribution that will be used in the analysis stage.       

\subsection{Run time}
In a typical run the IDL version of \PHI\ requires between $10^{4}$ to $10^{4.5}$ iterations for three simultaneously running chains to converge. The median total generation time for a $250 \times 250$ pixel image is $t_{total} = 0.029s$ for a single S\'{e}rsic model and $t_{total} =  0.041s$ for a S\'{e}rsic + exponential model. The wall clock time for a complete run on a 2.5 GHz Intel core i5 CPU is $\sim 10$ minutes and $\sim 20$ minutes for a single S\'{e}rsic and a S\'{e}rsic + exponential model, respectively.  Run times are similar for real and mock galaxies. The quoted times are for convolution with a $50 \times 50 $ pixel PSF.

\begin{figure}
\centering 
\includegraphics[width=0.5\textwidth, page=1, trim=40 140 60 180, clip]{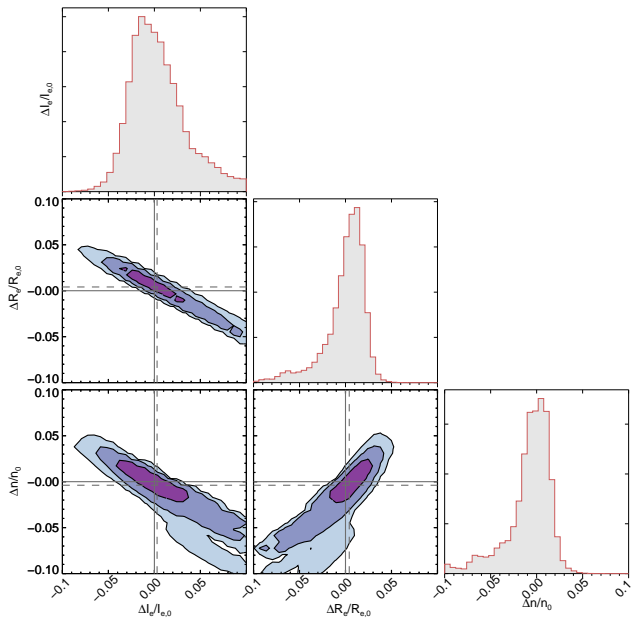}
\caption{The posterior fractional error distribution for the entire ensemble of synthetic elliptical galaxies. The marginalised error distributions for each structural parameter are shown on the diagonal. Joint marginal error distributions are shown in the off-diagonal panels. The colour contours represent the 68, 95, and 99\% confidence levels. The solid grey lines indicate the true values and the dashed lines represent the median of the posterior error distributions. A positive (negative) value indicates the fitted value is larger (smaller) than the true value.}
\label{fig:totbias_ell}
\end{figure}

\section{Application to synthetic galaxy images}\label{sec:synthetic}

In this section we use synthetic galaxy images to test the accuracy and robustness of \PHI. Synthetic galaxies lack the complexity present in real galaxies, but allow us to check for any systematic errors inherent in the method. The images were made to mimick Sloan Digital Sky Survey \citep[SDSS,][]{Str02} $i$-band  images as closely as possible, with a pixel scale of 0.396 arcsec/pixel, CCD gain of 4.86$e^−$/ADU and read-out noise of 5.76$e^−$. The appropriate noise level was estimated from the SDSS data frames by removing all objects and fitting a Poisson distribution to the residual counts. 

Fig.~\ref{fig:postone} shows the posterior distribution produced by \PHI\ for fitting an example two-component synthetic galaxy image with an inner S\'{e}rsic and outer exponential component as described in Section \ref{sec:model} and with priors as described in Section \ref{sec:priors}. The medians of the posterior distributions have a fractional error of at most 2\% in relation to the true values. We also clearly see that there is a strong covariance between parameters within the individual components, i.e. $I_{e}$, $R_{e}$, and $n$ for the S\'{e}rsic profile and $I_{0}$ and $h$ for the exponential component. There is also some covariance between the two components i.e. $n$ vs. $h$. The quantification of these covariances is important, as it may cause correlations between physical parameters (e.g. scaling relations) to appear stronger than they are in reality. The entire posterior distribution for a galaxy can be used when testing hypotheses about galaxy populations and this will be explored in a subsequent paper. 

\subsection{Population of synthetic galaxies}

\begin{figure*}
\centering 
\includegraphics[width=1.\textwidth, page=1, trim=40 150 30 180, clip]{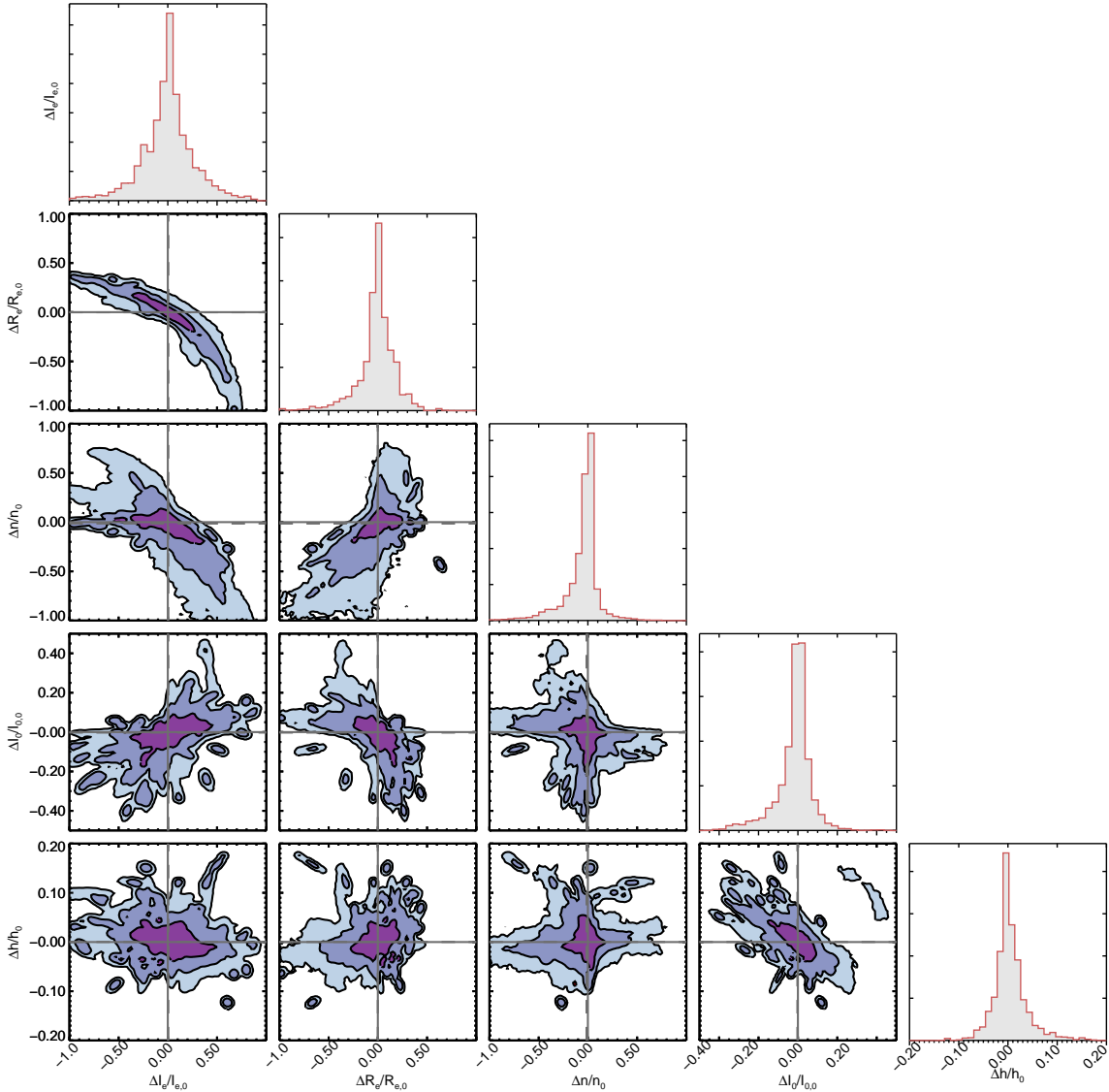}
\caption{Same as Fig.~\ref{fig:totbias_ell} for the synthetic two-component bulge+disc galaxies. }
\label{fig:totbias}
\end{figure*}

\begin{table}
\centering
\caption{Ranges for structural and geometric parameters used to simulate synthetic elliptical and bulge+disc galaxies. Geometric parameters are independently determined for each component.}\label{tab:synthetic}
\begin{tabular}{ll}
\hline
\textbf{Synthetic Elliptical }             &                                       \\
                                                          &                                       \\
\textbf{Parameter}                           & \textbf{{[}min, max{]}}  \\ \hline
S\'{e}rsic magnitude ($m_{S, i}$)         & {[}14, 17{]}                       \\
Effective radius ($R_{e}$ in arcsecs)   & {[}1.5, 6{]}                       \\
S\'{e}rsic index ($n$)                         & {[}1.9, 7.5{]}                     \\
                                                          &                                       \\ \hline
\textbf{Synthetic Bulge + Disc }      &                                       \\
                                                          &                                       \\
\textbf{Parameter}                          & \textbf{{[}min, max{]}}  \\ \hline
S\'{e}rsic magnitude ($m_{S, i}$)         & {[}15, 21{]}                     \\
Effective radius ($R_{e}$ in arcsecs)  & {[}0.4, 2.24{]}                  \\
S\'{e}rsic index ($n$)                        & {[}0.5, 7{]}                       \\
Exponential magnitude ($m_{E, i}$)   & {[}15, 18{]}                       \\
Scale length ($h$ in arcsecs)            & {[}1.3, 7{]}                       \\
                                                        &                                        \\ \hline
\textbf{Geometric parameters of bulge or disc}     &                                       \\
                                                        &                                       \\
\textbf{Parameter}                         & \textbf{{[}min, max{]}}  \\ \hline
Axial ratio ($q$)                               & {[}0.6, 1{]}                       \\
Position angle ($\textrm{PA}$ in degrees)   & {[}-360, 360{]}                
\end{tabular}
\end{table}

\begin{table}
\centering
\caption{The 16th, 50th and 84th percentiles of the posterior fractional error distributions for the synthetic elliptical and bulge+disc galaxies.}
\label{tab:bias}
\begin{tabular}{llll}
\hline
\textbf{Synthetic Elliptical } &                 &                        &                        \\
                                                 &                 &                        &                        \\
\textbf{Parameter}                               & \textbf{Median} & \textbf{16\% } & \textbf{84\% } \\ \hline
$\Delta I_{e}  / I_{e,in}$             & 0.003           & -0.022                 & 0.049                  \\
$\Delta R_{e} / R_{e,in}$             & 0.004           & -0.021                 & 0.018                  \\
$\Delta n / n_{in}$                        & -0.004          & -0.033                 & 0.012                  \\
                                                 &                 &                        &                        \\ \hline
\textbf{Synthetic Bulge + Disc } &                 &                        &                        \\
                                                 &                 &                        &                        \\
\textbf{Parameter}                               & \textbf{Median} & \textbf{16\% } & \textbf{84\% } \\ \hline
$\Delta I_{e} / I_{in}$                & 0.005           & -0.25                  & 0.22                   \\
$\Delta R_{e} / R_{e,in}$             & -0.001          & -0.15                  & 0.14                   \\
$\Delta n / n_{in}$                        & -0.014          & -0.19                  & 0.05                   \\
$\Delta I_{0} / I_{0,in}$             & -0.002          & -0.09                  & 0.04                   \\
$\Delta h / h_{in}$                        & 0.0001          & -0.02                  & 0.03                  
\end{tabular}
\end{table}

To test \PHI\ on a realistic range of galaxy structural parameters, we take the parameters of 260 elliptical and 380 bulge+disc galaxies in the SDSS fitted by  \citet{Gad09} with single S\'{e}rsic and S\'{e}rsic + exponential profiles respectively. Further details about this sample are given in Section \ref{sec:data}. Table \ref{tab:synthetic} summarises the range of parameters tested in this way. In particular, we test a large range in the bulge-to-total flux ratio of bulge+disc galaxies with $0.01<B/T< 0.8$. For each synthetic galaxy we also assume a Gaussian PSF with FWHM as provided by \citet{Gad09}. According to this paper, these values were largely taken from the SDSS DR2 imaging headers, although erroneous values were corrected for by fitting nearby stars. The mean FWHM of the sample is 1.5\arcsec, with a range between 0.4 and 2.6\arcsec. 

\subsection{Fractional errors on structural parameters}

To visualise the errors over a large sample of galaxies, we calculate the fractional error distribution on each parameter as $(x_{out} - x_{in}) / x_{in}$, where $x_{in}$ is the input parameter value and $x_{out}$ are the values given in the MCMC output. The stacked posterior fractional error distributions of all the synthetic galaxies are shown in Fig.~\ref{fig:totbias_ell} for elliptical galaxies and Fig.~\ref{fig:totbias} for bulge+disc galaxies. The clear covariances between the model parameters seen in Fig.~\ref{fig:postone} lead to a high degree of correlation in the fractional errors. These figures clearly show the well known degeneracies inherent in S\'{e}rsic fits: (i) if the effective intensity is overestimated (underestimated) the effective radius will be underestimated (overestimated) to compensate; (2) if the effective radius is overestimated (underestimated) the S\'{e}rsic index will also be overestimated (underestimated) to increase the concentration of intensity within a now larger effective radius.

Table \ref{tab:bias} presents the statistics of the posterior fractional error distributions. It can be seen that the systematic bias on parameters is minimal for both elliptical and bulge+disc galaxies, and the $1\sigma$ errors (16th and 84th percentiles) are usually $<20$\%. In the following subsections we look in more detail at the impact of degeneracies between parameters on obtaining unbiased estimates for the structural parameters and $B/T$. 

\subsubsection{Impact on bulge+disc structural parameters}\label{sec:BulgeBias}
\begin{figure*}
\centering 
\includegraphics[width=1. \textwidth, page=1, trim=0 120 50 100, clip]{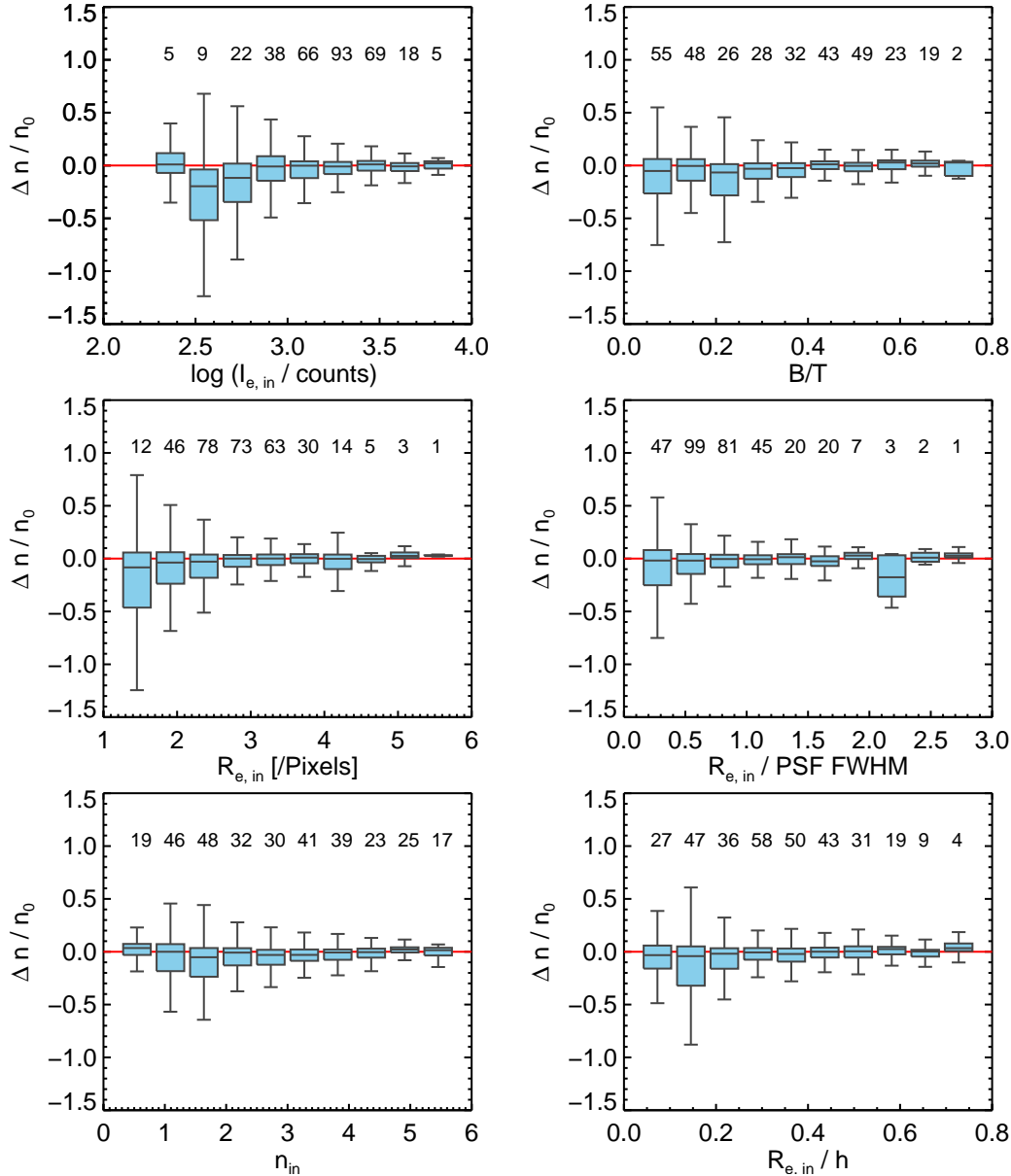}
\caption{Box-plots representing the fractional error on $n$ for bulge+disc synthetic galaxies as a function of input parameters: $I_{e}$, $R_{e}$, $n$, $B/T$, $R_{e}/h$, as well as the ratio between input $R_{e}$ and the PSF FWHM. The box limits represent the 16th, and the 84th percentiles of the posterior fractional error distributions and the horizontal line shows the 50th percentile. The whiskers show the total extent of the distributions. The value above each box gives the number of galaxies in each bin. }
\label{fig:onn}
\end{figure*}

\begin{figure*}
\centering 
\includegraphics[width=0.48 \textwidth, page=2, trim=45 145 110 250, clip]{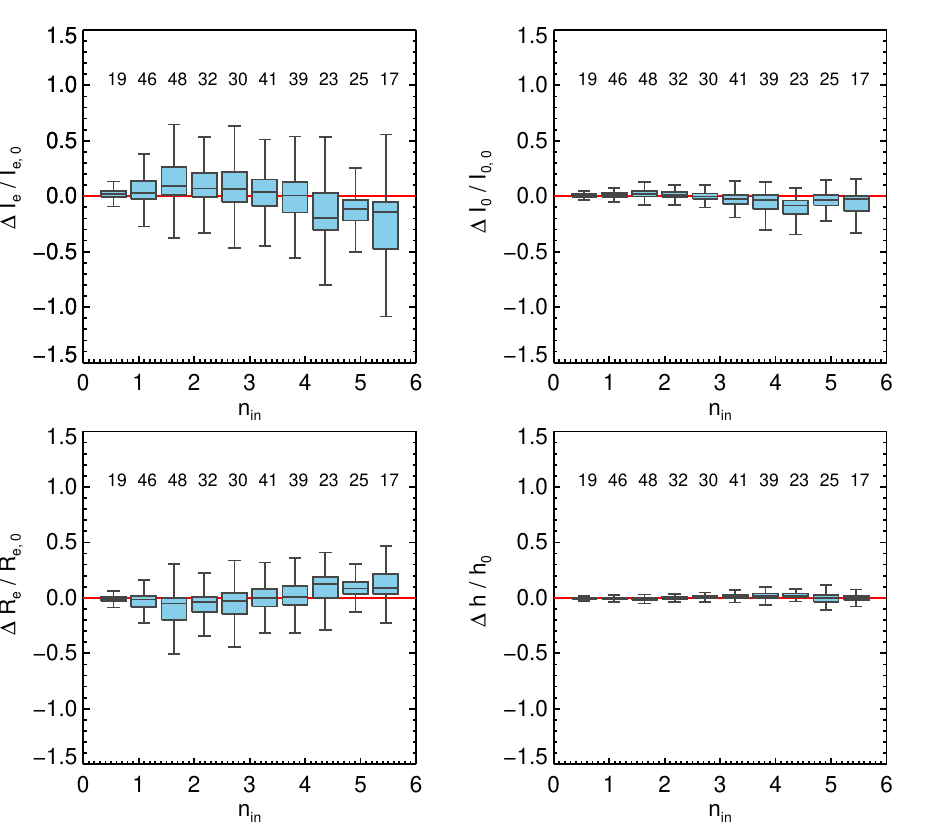}
\includegraphics[width=0.48 \textwidth, page=3, trim=45 145 110 250, clip]{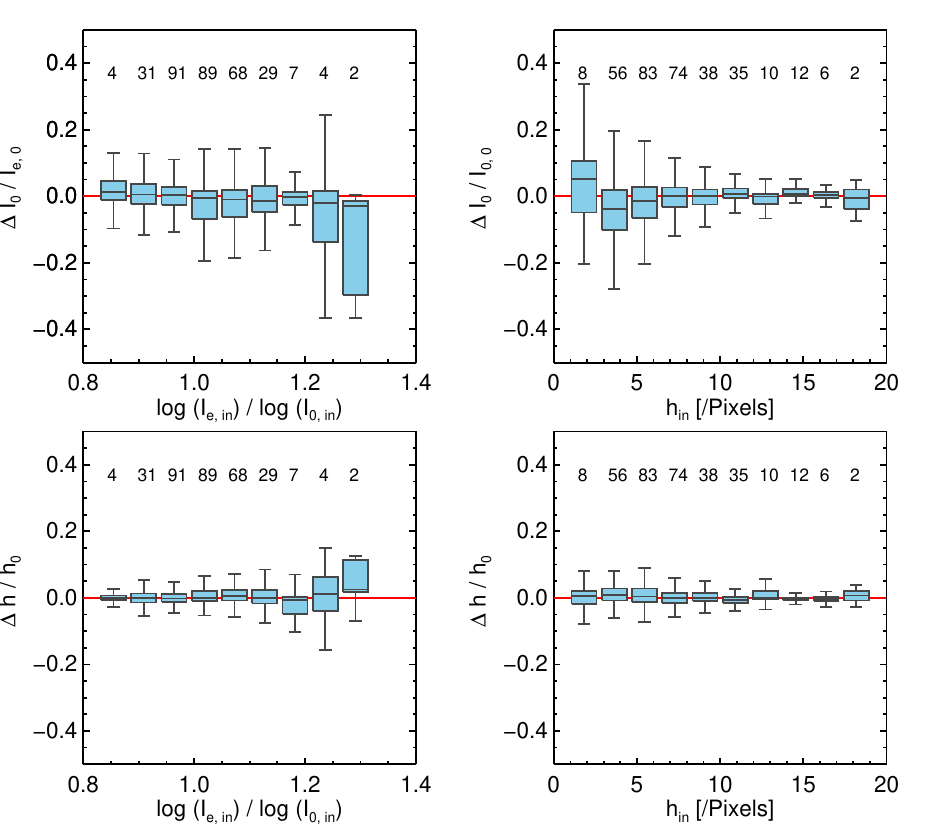}
\caption{The fractional errors on $I_{e}$,  $I_{0}$, $R_{e}$ and $h$ as a function of input $n$ {\it (Left)} and  $I_{0}$ and $h$ as a function of $I_e/I_0$ and $h$ {\it (Right)}. See the caption of Fig.~\ref{fig:onn} for more details.}
\label{fig:offn}
\end{figure*}

Fig.~\ref{fig:onn} shows the fractional error on $n$ as a function of a selection of structural parameters for the two component bulge+disc synthetic galaxies, as well as the ratio between $R_e$ and the PSF FWHM.  The blue box indicates the 16th and 84th percentiles of the marginalised posterior fractional error distribution, with the horizontal line showing the median, and the whiskers indicating the maximum extent of the posteriors.  As we saw in Fig~\ref{fig:totbias} the fractional errors are typically $<20$\% and the median shows no overall bias. However, this figure shows how the errors on $n$ increase for models with smaller $I_e$, smaller $B/T$, smaller $R_e$, smaller ratio of $R_e$ to PSF FWHM and smaller $R_e/h$.  The errors appear to be largest for $n\sim2$ bulges. 

Similarly, the two left-hand panels of Fig.~\ref{fig:offn} show the fractional error on the other structural parameters as a function of input $n$. As before we see that there is typically no overall bias in the estimation of parameters, with the median sitting close to zero. The exception is at large $n$ where a bias of $\sim12$\% and $\sim10$\% is found on $R_e$ and $I_e$ respectively for $n=6$. For high values of S\'{e}rsic index the algorithm also takes longer to converge, suggesting difficulties in this region. Further investigation suggests these difficulties are caused by a flattening in the likelihood space for galaxies with higher $n$ values ($n \geq 4.5$). Subsequent changes in the $n > 4.5$ region therefore result in little variation in the likelihood values, thus decreasing the accuracy and precision.  We explore this specific case of high-$n$ bulges in more detail in Appendix \ref{sec:app}, finding that they are particularly susceptible to strong degeneracies between $R_e$ and $n$ when $B/T$ is low. 

\subsubsection{Impact on $B/T$}

\begin{figure*}
\centering 
\includegraphics[width=1\textwidth, page=5, trim=0 120 50 100, clip]{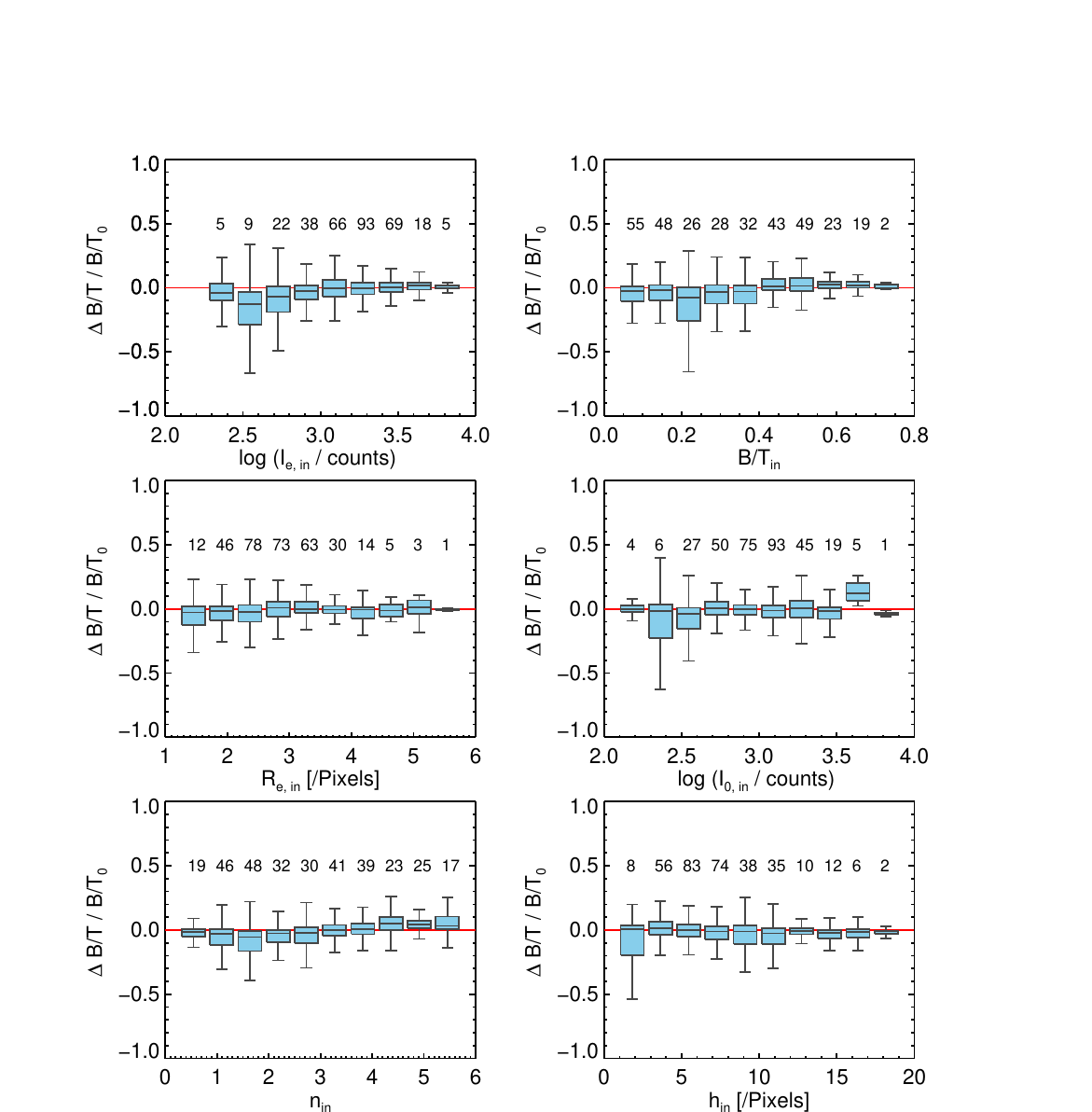}
\caption{Same as Fig.~\ref{fig:onn} but showing the fractional errors on $B/T$ as a function of input $I_{e}$, $B/T$, $R_{e}$,  $I_{0}$, $n$,  and $h$.}
\label{fig:btbi}
\end{figure*}

Galaxy morphology is commonly parametrised by the bulge-to-total flux ratio ($B/T$), it is therefore important to assess the biases on this particular parameter caused by correlations in the parameter errors observed in Fig.~\ref{fig:totbias}. Fig.~\ref{fig:btbi} shows the fractional error on the measured $B/T$ as a function of input parameters $I_{e}$, $B/T$, $R_{e}$,  $I_{0}$, $n$ and $h$. The median values are remarkably stable for the whole range of models tested here, indicating that $B/T$ should provide a robust parametrisation of galaxy morphology. 

\subsubsection{Impact on the disc parameters}

The two right-hand panels of Fig.~\ref{fig:offn} show the fractional errors on the disc structural parameters $I_{0}$ and $h$, as a function of input $I_e/I_0$ and $h$. Additionally, the center-left column of Fig.~\ref{fig:offn} shows how the S\'{e}rsic index affects for the same parameters. The fractional errors on the disc scale length $h$ are the smallest of all structural parameters, this is because the disc is well resolved in the images. There is no clear bias in the measured disc parameters, although as expected the fractional errors increase where the disc becomes less dominant (as traced by the $I_e/I_0$ ratio) and the disc scale length becomes smaller. 

\section{Application to data}\label{sec:data}
\begin{figure*}
\centering 
\includegraphics[width=1 \textwidth, page=1, trim=50 370 30 100, clip]{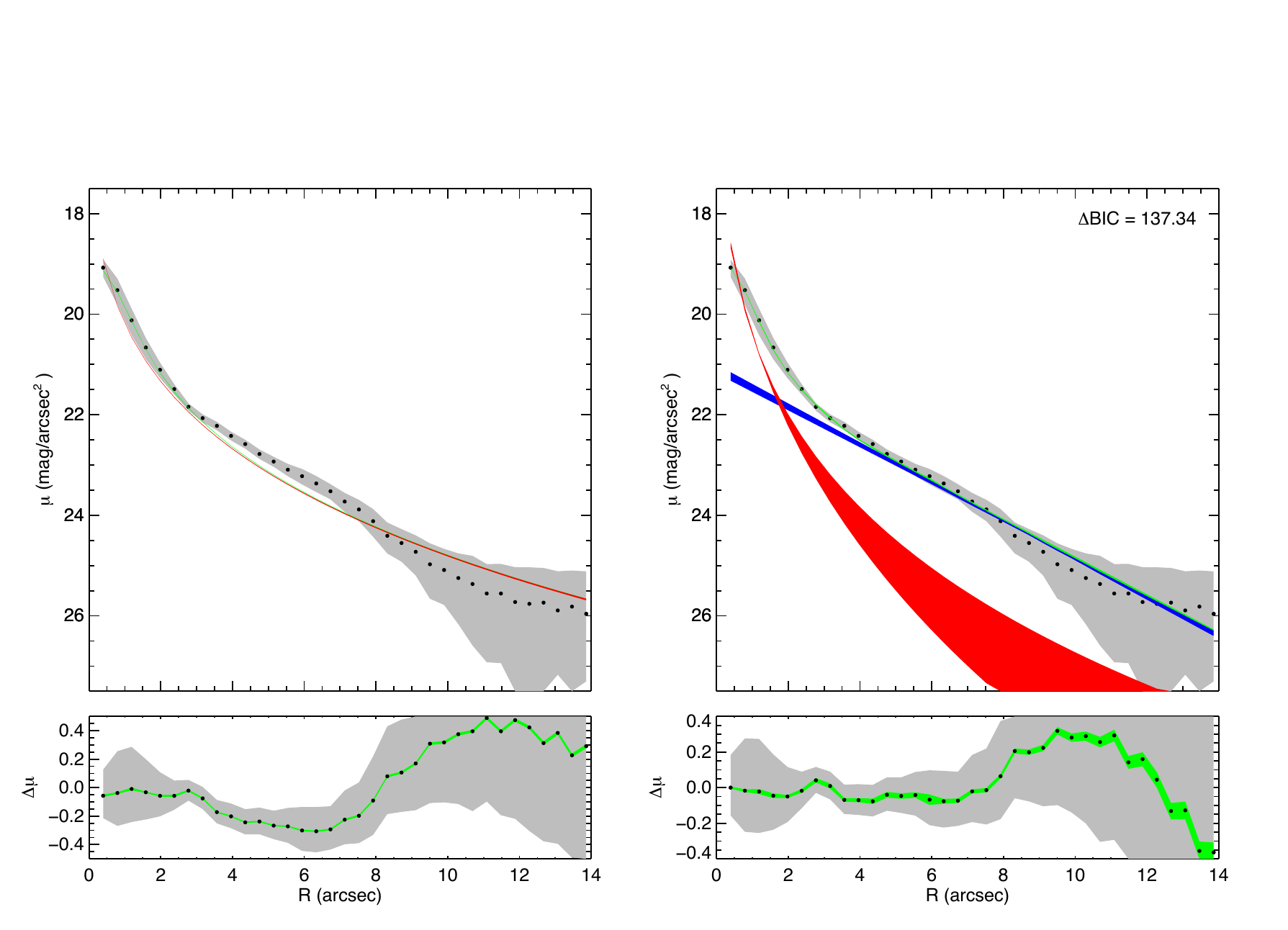}
\caption{Ellipse-averaged radial profile of the surface brightness of SDSS galaxy  J084149.16+504711.1  (black dots) with the root mean square error of the flux values in a given ellipse (grey region). The green band shows the synthetic galaxies generated from random draws from the posterior parameter distributions output by \PHI, with the resulting models convolved with the PSF.  The lower panels show the residuals between the data and the model created with the posterior median parameters. {\it Left:} for a one-component S\'{e}rsic model. {\it Right:} for a two-component S\'{e}rsic+exponential model. The blue and red bands show the random draws for the exponential and  S\'{e}rsic components respectively, without PSF convolution. The $\Delta\textrm{BIC}$ for this galaxy is $137$, indicating that an improved fit is provided by the two-component model (see Section \ref{sec:BIC}).}
\label{fig:pred}
\end{figure*}

\begin{figure*}
\centering 
\includegraphics[width=1 \textwidth, page=1, trim=170 150 170 100, clip]{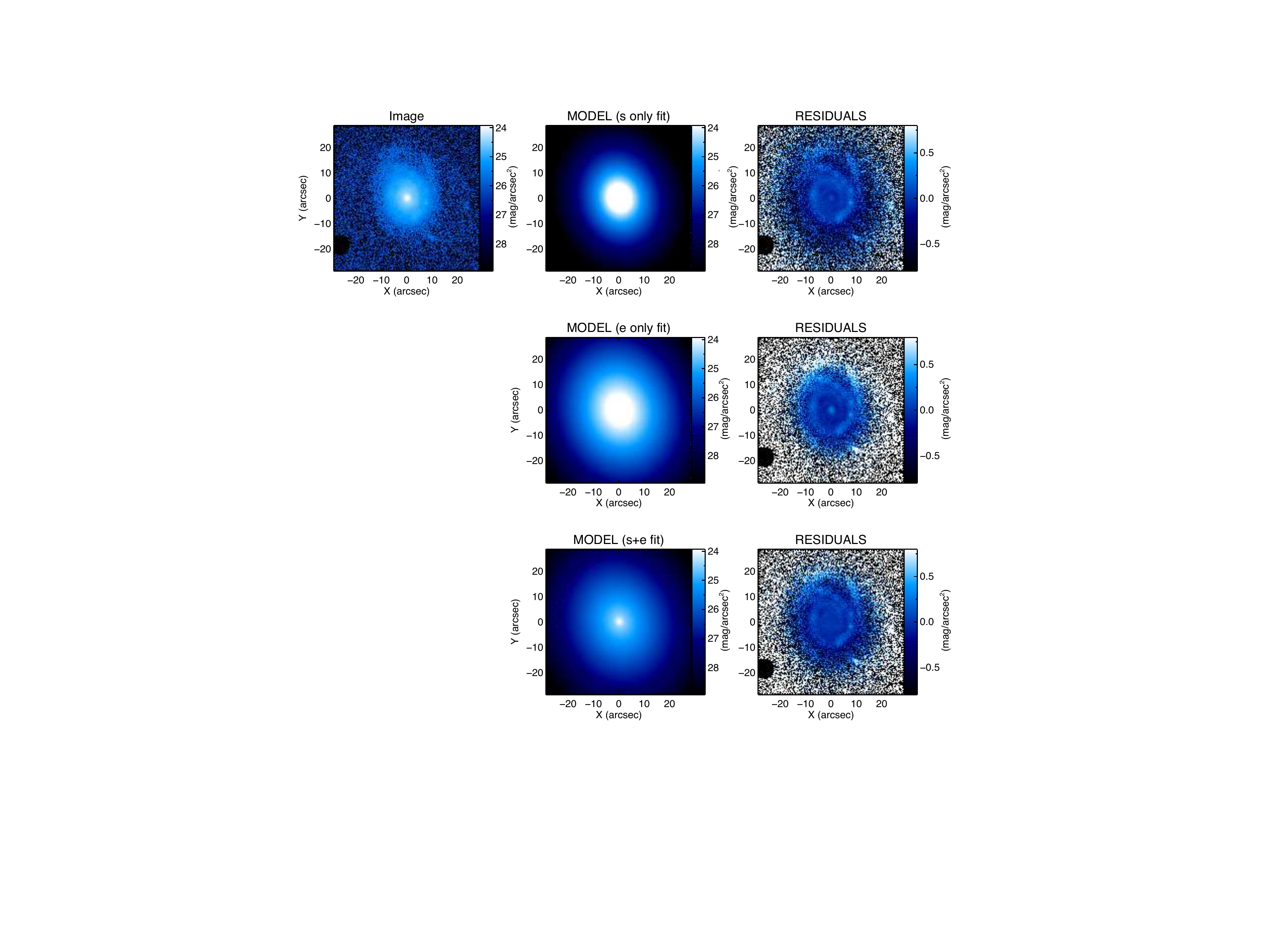}
\caption{A 2D representation for the SDSS galaxy presented in Fig.~\ref{fig:pred}. The top row shows the data (left), the S\'{e}rsic-only model fit (middle) and the residual (right). The second row shows an exponential only fit (i.e. with fixed $n=1$) and corresponding residual. The bottom row shows the bulge-disc model fit and corresponding residual. All the models were made using the medians from the posterior distributions.}
\label{fig:Example2d}
\end{figure*}

The next step is to apply \PHI\ to a sample of real galaxy images to assess the functionality and robustness of the method in a fully realistic scenario. It is first useful to visualise the fits from \PHI. To do this, we collapse the 2D data and model images into one-dimension (1D) using ellipse averaging. In order to represent the full posterior uncertainty on the fitted parameters, and therefore the profiles, we draw randomly from the parameter posterior distributions to estimate the median surface brightness and 1$\sigma$ errors as a function of radius. 

Fig.~\ref{fig:pred} shows the 1D surface brightness profile of a bulge+disc galaxy in the \citet{Gad09} sample to which we have fitted both a one-component  S\'{e}rsic model (left panel), and a two-component S\'{e}rsic+exponential model (right panel). The black dots and grey region shows the image data and root-mean square of the flux values in a given ellipse. The green band shows the ensemble of model fits generated from drawing parameters randomly from the posterior distribution and convolving the resulting model with the PSF. Residuals from the model generated from the median posterior parameters are shown in the lower panels. In the case of the one-component model, the central region is well fitted, but the model deviates significantly from the data beyond $3\arcsec$. This is improved by the two-component model, with an acceptable fit out to $8\arcsec$, although the residuals beyond this radius suggest that this galaxy has a truncated disc. 

Fig.~\ref{fig:Example2d} shows the 2D surface brightness profile of the same bulge+disc galaxy presented in Fig.~\ref{fig:pred}. The top row shows the image, the fitted one-component S\'{e}rsic model and the residual. The middle and bottom rows show an exponential only and the S\'{e}rsic+exponential models respectively, with their residuals. The exponential only model is a one-component S\'{e}rsic model with fixed $n = 1$; this model is not used in this paper, but presented here to aid understanding. The models were generated from the posterior medians from the MCMC outputs. As in Fig.~\ref{fig:pred} we see that the one-component S\'{e}rsic model fits the central region well, which can not be fit with a pure $n=1$ profile (middle planel).  The two-component model provides the lowest residuals in both the central and outer regions, leaving a clear signal from the spiral arms that are not included in the model.         

Although the graphical and visual representation are useful tools to ensure the code is working as it should, the quantitative model comparison presented below in Section \ref{sec:BIC} is required in order to make statistical claims about which model is best.

\subsection{Analysis of SDSS images}

We study SDSS $i$-band images of galaxies with stellar masses $>10^{10} M_{\odot}$, $0.04 \leq z \leq 0.06$ and $q \geq 0.9$ that were previously analysed by \citet{Gad09} using the BUDDA code \citep{Des04} to perform bulge/disc/bar 2D photometric decompositions. \citet{Gad09} selected 1000 galaxies and separated them into elliptical and bulge+disc based on the $i$-band Petrosian concentration index, $C$, as given in the SDSS database, defining ellipticals to have $C > 3$, disc galaxies to have $C < 2.5$ and bulge+disc galaxies to lie in between. Despite the constraints imposed on axis ratio and concentration parameter, the sample is considered to be a fair representation of the galaxy population in the local Universe. We select only those galaxies that were classified as elliptical or bulge+disc, and remove from the sample any galaxies that were found to have a bar (visually identified from residual fits). As noted above, for this code presentation paper we have chosen to focus on one- and two-component galaxies, although the code is able to fit any model specified by the user. Barred galaxies will be studied with \PHI\ in a future publication. This leaves us with 260 elliptical and 380 bulge+disc galaxies.  

In \citet{Gad09} the imaging used to classify the sample and perform the 2D photometric decompositions was from the SDSS data release 2 \citep[DR2][]{Aba04}. For this study we use the SDSS DR7 \citep{Aba04} images. Moffat PSFs were obtained for each individual galaxy by fitting a Moffat function to 5-10 stars nearby to each galaxy and using the median value for each parameter. Segmentation maps were created following a similar approach to that used by Source Extractor \citep{Bert96}. There is no concern about overlapping sources as the galaxies were originally selected to be isolated. We took the gain, readnoise and sky values from the SDSS image headers, and combined these with the galaxy shot noise to compute the weight maps in the standard way.

\PHI\ was run on a square cut-out image, typically $250\times250$ pixels in size, following removal of unwanted sources using the segmentation map. The size of the cut-out was selected to include the entire galaxy, although slightly smaller cut-outs were used in a few cases where the galaxy fell close to the edge of the image. We ran \PHI\ with 3 simultaneous chains for both one-component  S\'{e}rsic and two-component S\'{e}rsic+exponential models on every galaxy present in both samples, in order to perform a model comparison in Section \ref{sec:BIC}. We additionally analysed the same images using GASP2D \citep[see ][]{Men14, Men17}, with the same PSF, weight maps and masks. Finally, we compared our results to those presented in \citet{Gad09} which use SDSS DR2 images, and primarily takes the PSF values from the SDSS image headers. GASP2D uses the LM and BUDDA uses the Nelder-Mead simplex minimisation method so we can investigate whether any differences that arise are due to different codes or different images/treatment of the images.

\subsection{Comparison of elliptical galaxies}
\begin{figure}
\centering 
\includegraphics[width=0.5 \textwidth, page=1, trim=50 330 250 100, clip]{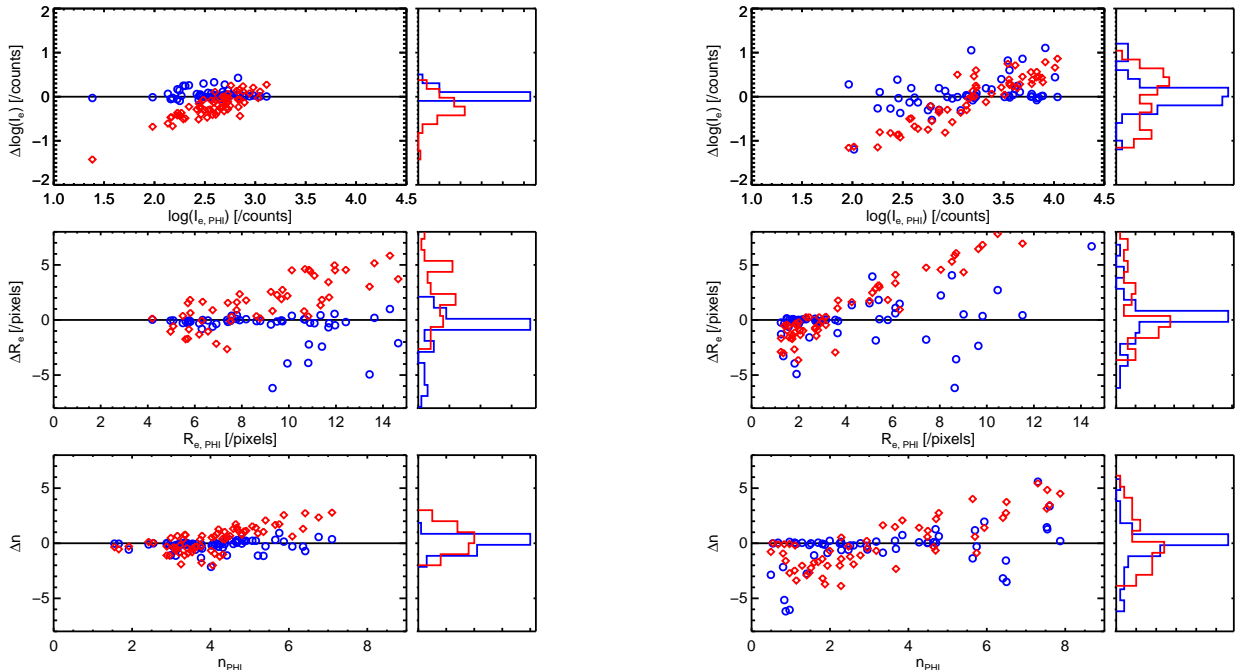}
\caption{The differences between parameter estimates for elliptical galaxy images fit with a one-component S\'{e}rsic profile. From top to bottom the parameters are: the effective intensity ($I_{e}$), effective radius ($R_{e}$) and S\'{e}rsic index ($n$). The blue circles and red diamonds are the posterior medians given by \PHI\ compared to the best-fit from GASP2D and \citet{Gad09} respectively (i.e. \PHI\ - GASP2D and \PHI\ - Gad09). The projected distributions are shown to the right, with the blue and red histogram comparing \PHI\ to  GASP2D and \citet{Gad09}, respectively.}
\label{fig:comp_ell}
\end{figure}

Fig.~\ref{fig:comp_ell} compares the measured S\'{e}rsic profile parameters for the galaxies classified as elliptical in \citet{Gad09}. The blue circles show the difference between \PHI\  and GASP2D and the red diamonds the difference between \PHI\ and \citet{Gad09}. We removed any catastrophic failures that occurred when running GASP2D, leaving 250 elliptical and 350 bulge+disc galaxies for the final sample.

We focus first on the comparison between \texttt{PHI}  and GASP2D, where the same images are fit with the same PSFs, weight maps, and segmentation maps. The results are well correlated and agree on average. The standard deviations for the parameter differences are $\sigma_{log(I_{e})} = 0.10$ counts, $\sigma_{R_{e}} = 2.0$ pixels and $\sigma_{n} = 0.49$. There is a subtle deviation from an exact one-to-one match at larger values of $R_{e}$ and $n$. In GASP2D, a run is determined to have reached the global minimum when the deviations between the $\chi^{2}$ of two consecutive iterations is lower than a given threshold. This threshold cut results in changing errors with parameter values: due to the exponential nature of the S\'{e}rsic profile,  changes to $n$ where $n\leq2$ have a greater impact on the surface brightness profile than at larger values of $n$. This effect is not seen in \PHI\ due to the efficient exploration of parameter space and adaptable step sizes. 

When we compare the median posterior parameter values measured by \PHI\ to the best-fit values obtained by \citet{Gad09} we see significant differences for all the parameters. Most notably, the values fit by \texttt{PHI} span larger ranges than found by \citet{Gad09}. For example, the distribution of $n$ in \citet{Gad09} has a mean of $3.8$ and standard deviation of $0.9$, compared to a mean of $4.1$ and standard deviation of $1.2$ found by \PHI. This difference in range results in the visible trend between the size of the offset and the value of the fitted parameter.

\subsection{Comparison of bulge+disc galaxies}
\begin{figure}
\centering 
\includegraphics[width=0.5 \textwidth, page=1, trim=50 120 250 100, clip]{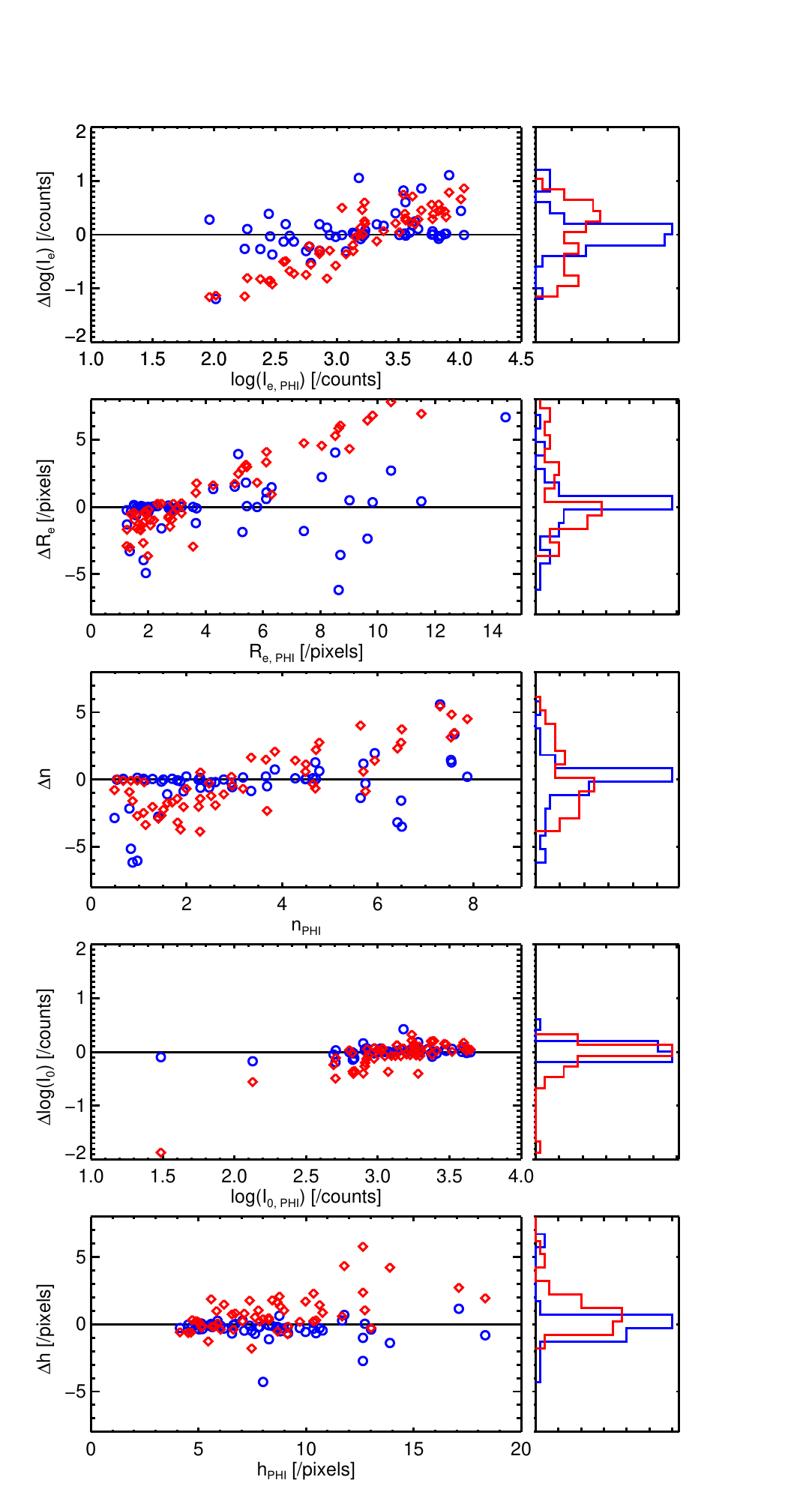}
\caption{The same as Fig.~\ref{fig:comp_ell} but for the sample of bulge+disc galaxies. From top to bottom we have the effective intensity ($I_{e}$), the effective radius ($R_{e}$), the S\'{e}rsic index ($n$), the central intensity ($I_{0}$) and the scale length ($h$).}
\label{Fig:comp_bd}
\end{figure}

 Fig.~\ref{Fig:comp_bd} compares the results between \PHI, GASP2D and \citet{Gad09} for the sample of bulge+disc galaxies. The results found by GASP2D and \PHI\ are broadly consistent, with no obvious bias. Standard deviations for the parameter differences are $\sigma_{log(I_{e})} = 0.4$ counts, $\sigma_{R_{e}} = 2.0$ pixels, $\sigma_{n} = 1.9$, $\sigma_{log(I_{0})} = 0.1$ counts, $\sigma_h =  1.5$ pixels. 

Comparing between the posterior median values found by \PHI\ and the best-fit values found by  \citet{Gad09} we see that the bulge components are systematically different, as a function of parameter value. This is identical to the pattern seen for the elliptical galaxies and is caused by the much smaller range of parameter values that are fitted by \citet{Gad09} compared to \PHI. On the other hand, the values found for the disc parameters are much more consistent, which agrees with our analysis of synthetic galaxies where disc parameters show much smaller fractional errors. It is clear from our analysis of both single S\'{e}rsic and S\'{e}rsic+Exponential galaxies with BUDDA, GASP2D and \texttt{PHI}, that significant disagreements appear in the fitted bulge parameters when different code, images (DR2 vs. DR7), weights and PSFs are used.  This highlights a fundamental limitation of bulge+disc decomposition, in that the estimation of bulge parameters will always be susceptible to biases and systematics when they are barely resolved in comparison to the PSF. We explore this issue with the synthetic galaxies in Section \ref{sec:BulgeBias} and Appendix \ref{sec:app}. Using a fully Bayesian code such as \PHI\ allows you to explore potential biases, errors and covariances with ease, but ultimately a full code comparison study is clearly required to understand the limitations in more detail.

\section{Model comparison}\label{sec:BIC}
\begin{figure}
\centering 
\includegraphics[width=0.6 \textwidth, page=7, trim=160 120 100 100, clip]{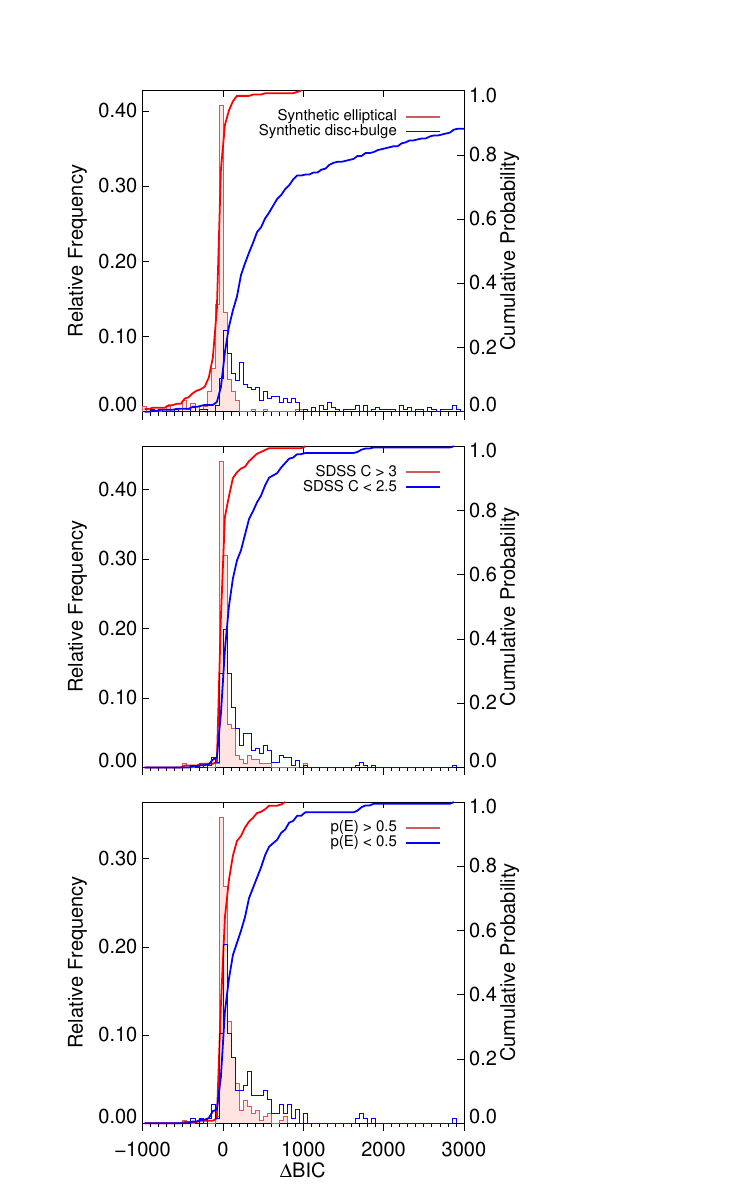}
\caption{Histograms and cumulative distributions showing the $\Delta\textrm{BIC} = \textrm{BIC}_{1c} - \textrm{BIC}_{2c}$ distributions. A positive value shows that the two-component model is preferred. {\it Top:} Synthetic one- and two-component  galaxies, shown as red and blue lines respectively; {\it Middle:} SDSS galaxies classified by their concentration index as elliptical (red line) or bulge+disc (blue line); {\it Bottom:} SDSS galaxies classified as elliptical (red line) or disky (blue line) by the machine learning algorithm of \citet{Hue11}. }
\label{fig:BIC}
\end{figure}

\begin{table}
\centering
\caption{The statistics of the $\Delta$BIC distributions for the synthetic and SDSS samples. Where $C$ is the $i$-band concentration parameter provided in the SDSS catalogue, and p(E) is the probability that the galaxy is an elliptical from the machine learning classification of \citet{Hue11}.}
\label{tab:BIC}
\begin{tabular}{lllll}
\hline
\textbf{Synthetic $\Delta$BIC}                         &         & \textbf{} &              &              \\
                                           &         &           &              &              \\
                                           & Mean    & Median    & 16\% & 84\% \\ \hline
\multicolumn{1}{l|}{Elliptical}            & -427.47 & -31.09    & -149.74      & 13.36        \\
\multicolumn{1}{l|}{Bulge + disc}          & 1218.51 & 341.64    & 35.13        & 2272.50      \\
                                           &         &           &              &              \\ \hline
\textbf{SDSS $\Delta$BIC}                              &         & \textbf{} &              &              \\
                                           &         &           &              &              \\
                                           & Mean    & Median    & 16\% & 84\% \\ \hline
\multicolumn{1}{l|}{C \textgreater 3}      & -89.70  & 2.78      & -25.30       & 89.63        \\
\multicolumn{1}{l|}{C \textless 2.5}       & 225.67  & 99.43     & -4.29        & 480.77       \\
\multicolumn{1}{l|}{p(E) \textgreater 0.5} & -16.88  & 18.82     & -21.91       & 152.45       \\
\multicolumn{1}{l|}{p(E) \textless 0.5}    & 280.88  & 134.56    & 3.69         & 551.71      
\end{tabular}
\end{table}

In this section we test the use of the $\Delta\textrm{BIC}$ introduced in Section \ref{sec:method} to formally distinguish between one and two-component galaxies. We define the $\Delta\textrm{BIC} = \textrm{BIC}_{1c}- \textrm{BIC}_{2c}$, so a larger value indicates that a two-component model is preferred. In simple and ideal situations a $\Delta\textrm{BIC}>10$ is typically taken to be decisive evidence that a more complex model is preferred over a simpler one. However, in the case of image decomposition simulations are required to inform the choice of boundaries. 

For every synthetic galaxy in Section \ref{sec:synthetic} we have performed a one- and two-component fit; the distributions in $\Delta\textrm{BIC}$ are shown in the top panel of Fig.~\ref{fig:BIC} with statistics of the distributions summarised in Table \ref{tab:BIC}.  The one-component/elliptical synthetic galaxies show a very tight $\Delta\textrm{BIC}$ distribution centred on zero, while the two component/bulge+disc synthetic galaxies have larger $\Delta\textrm{BIC}$ values. The results from the synthetic galaxies suggest that a minimum $\Delta\textrm{BIC}$ value of 13.4 could be used to differentiate one and two-component galaxies. This would incorrectly classify 16\% of one-component galaxies as two-component (i.e. contaminant level), but would identify 82\% of two-component galaxies correctly (i.e. high completeness). 

The middle panel shows the $\Delta\textrm{BIC}$ distribution for the galaxies studied in Section \ref{sec:data} and classified by their concentration index as either elliptical or bulge+disc. Table \ref{tab:BIC} shows that the low concentration index galaxies have a significantly higher median $\Delta\textrm{BIC} = 99.43$ than the high concentration index galaxies with $\Delta\textrm{BIC} = 2.78$. However, there is no clear differentiating line between the two samples. 22\% of galaxies with $C>3$ are classified as 2-component systems by this method, and 38\% of galaxies with $C<2.5$ are classified as one-component. This shows that classifying galaxies by concentration index is not equivalent to classifying them by the results of 2D photometric bulge-disc decomposition. 

Finally, we compare with a machine learning morphological classification method by \citet{Hue11}, based on support vector machines. They assign a probability to each galaxy that it is an elliptical, S0, SAB or SCD. The algorithm was trained on visual classifications from the Galaxy Zoo first release catalogue \citep{Lin08,Lin11}. We classify as elliptical any galaxy with  $p(E)>0.5$, and plot the distributions of $\Delta\textrm{BIC}$ in the lower panel of Fig.~\ref{fig:BIC} for galaxies above and below this cut. We see a similar result to the case of classification by concentration index, with galaxies with higher $\Delta\textrm{BIC}$ more likely to have a disc, however, there is no clean demarcation between the two samples. 36\% of galaxies with $p(E)>0.5$ are classified as 2-component systems by their $\Delta\textrm{BIC}$, and 38\% of galaxies with $p(E)<0.5$ are classified as one-component. We note that this will include one-component disc galaxies, so their classification may in fact agree if we were to look in detail at the fitted parameters. 

In the synthetic galaxies the $\Delta\textrm{BIC}$ can clearly be used as a classification method to separate one- and two-component galaxies. However, for real galaxies the lack of any significant demarcation between galaxies classified by other methods suggests that the complex structure of real galaxies limits the usefulness of the  $\Delta\textrm{BIC}$ approach, certainly for a simple bulge+disc model as studied here. While galaxies with higher values of $\Delta\textrm{BIC}$ will have a higher probability of having a disc, we advocate that the $\Delta\textrm{BIC}$ should be used in combination with other methods to determine the number of structural components in a galaxy.

\section{Summary}\label{sec:summary}
We have used a new fitting algorithm (\texttt{PHI}) to perform 2D photometric decompositions of galaxy images from a Bayesian perspective. \PHI\ offers a number of significant advantages for estimating surface brightness profile parameters over traditional downhill optimisation algorithms:
 \begin{enumerate}
 \renewcommand{\labelenumi}{\Roman{enumi}.}  
 
   \item \PHI\ uses a triple layer approach to effectively and efficiently explore the complex parameter space. The first layer uses a blocked adaptive Metropolis algorithm to obtain an estimate of the scale for each parameter in the chain. The second layer uses an adaptive Metropolis algorithm with the purpose of estimating the target covariance matrix. The final level uses this estimated covariance matrix to quickly and effectively explore the parameter space. This reduces the chances of local minima trapping.     
   
   \item The algorithm naturally and explicitly incorporates priors that force the parameters to be realistic and physical, e.g., positive in the case of the dimensions and intensities. These priors replace the need for filtering processes to remove non-physical parameter outcomes. 
   
   \item Priors on parameters can be combined to further strengthen the model in an explicit way. In this paper, to prevent the reversal of components (\textit{i.e.} the desired inner component profile switching to fit the outer and vice versa) we use a prior combination that specifies that the bulges of galaxies are better modelled by a S\'{e}rsic profile and the discs are described by an exponential profile. We do this via a Newton-Raphson algorithm to determine the crossing points in the total light profile, as well as ensuring that the bulge is the dominant component in the central region. 
   
  \item \PHI\ gives the full posterior probability distribution for a set of model parameters. This is a powerful description of the model uncertainties that can be used in further analyses of galaxy structures.  

 \end{enumerate}

We used a sample of synthetic galaxies with SDSS-like image properties to ensure that there are no internal systematics due to the code, and to investigate the effect of galaxy properties on our ability to recover unbiased and well constrained structural parameters. In bulge+disc galaxies we find that the bulge structural parameters are recovered less well than those of the disc, particularly when the bulge contributes a lower fraction to the luminosity, or is barely resolved with respect to the pixel scale or PSF. The only systematic biases occur in bulge+disc galaxies with high bulge S\'{e}rsic index ($n>5$), where the code fits a bulge with an effective radius that is too large by 50\% and a central intensity that is too small by 20\%. No bias is found in the bulge-to-total luminosity ratio, which is important given the popularity of this quantity for parametrising galaxy morphologies. 

We have also applied \texttt{PHI}  to a sample of SDSS galaxies to compare with previous algorithms. Under the same image conditions, \textit{i.e.} images with the same masks, weights and PSF, \texttt{PHI} achieves consistent results with a standard minimisation code, with a low level of scatter. This validates both algorithms and approaches when assessing galaxy structures in the nearby Universe. However, we found less consistency when comparing to results from a previous analysis performed on different images, with different image conditions. The bulge structural parameters were the most affected, which we believe is due to the limited resolving power of SDSS images for local galaxy bulges. 

Finally, we investigated the Bayesian Information Criterion (BIC) as a method for deciding whether a galaxy has one- and two-components. In synthetic images the BIC cleanly separates the two types of galaxies, however, for real galaxies there was a less clean demarcation between galaxies classified as elliptical or bulge+disc by other methods in the literature. This suggests that the complexities of real galaxies are preventing a clean statistical test, and the BIC may need to be used in tandem with other methods to ensure that the correct model is selected. 

For future large-area, deep optical surveys such as Euclid and LSST, a full Bayesian analysis of local galaxy morphologies will be essential for unlocking the remaining unanswered questions about galaxy structures. Both fast, non-parametric approaches and full Bayesian fitting methods will need to be employed to quantify galaxy structures and successfully link them to the assembly history of galaxies. In the era of massive cosmological simulations taking galaxy evolution into a quantitative comparative science, observers must be careful to account for degeneracies between structural parameters when scaling relations are calculated.The addition of \texttt{PHI} into the 2D photometric decomposition toolbox will help advance our future understanding of galaxy properties. 

\section*{Acknowledgements}

We  thank the  anonymous referee  for her/his  many valuable  comments
which helped to improve this  paper. JJA, JMA, and VW acknowledge support of the European
Research Council via the award of a starting grant (SEDMorph;
P.I. V. Wild).  JMA acknowledge support
from the Spanish  Ministerio de Economia y  Competitividad (MINECO) by
the grant AYA2013-43188-P.

\bibliographystyle{mn2e}
\bibliography{litbib_chp3}{}

\appendix
\section{Limitations to modeling galaxy bulges}\label{sec:app}

Here we use a library of synthetic galaxies to investigate how the final parameter distributions are dependent on the resolution of the components, \textit{i.e.}, the total number of pixels that make up each component. It is clear that the ability to resolve the inner component (the bulge) will diminish as $R_{e}$ tends to the Nyquist limit. A decrease in resolution will effect the precision of other parameters of interest, which we demonstrate here. For these synthetic galaxies we fix $h = 10$ pixels as well as keeping $I_{0} = 10^{2.8}$ counts to ensure the disc is well resolved. These values are typical for the SDSS sample studied in Section \ref{sec:data}.

\begin{figure*}
\centering 
\includegraphics[width=1 \textwidth, trim=0 130 0 100, clip]{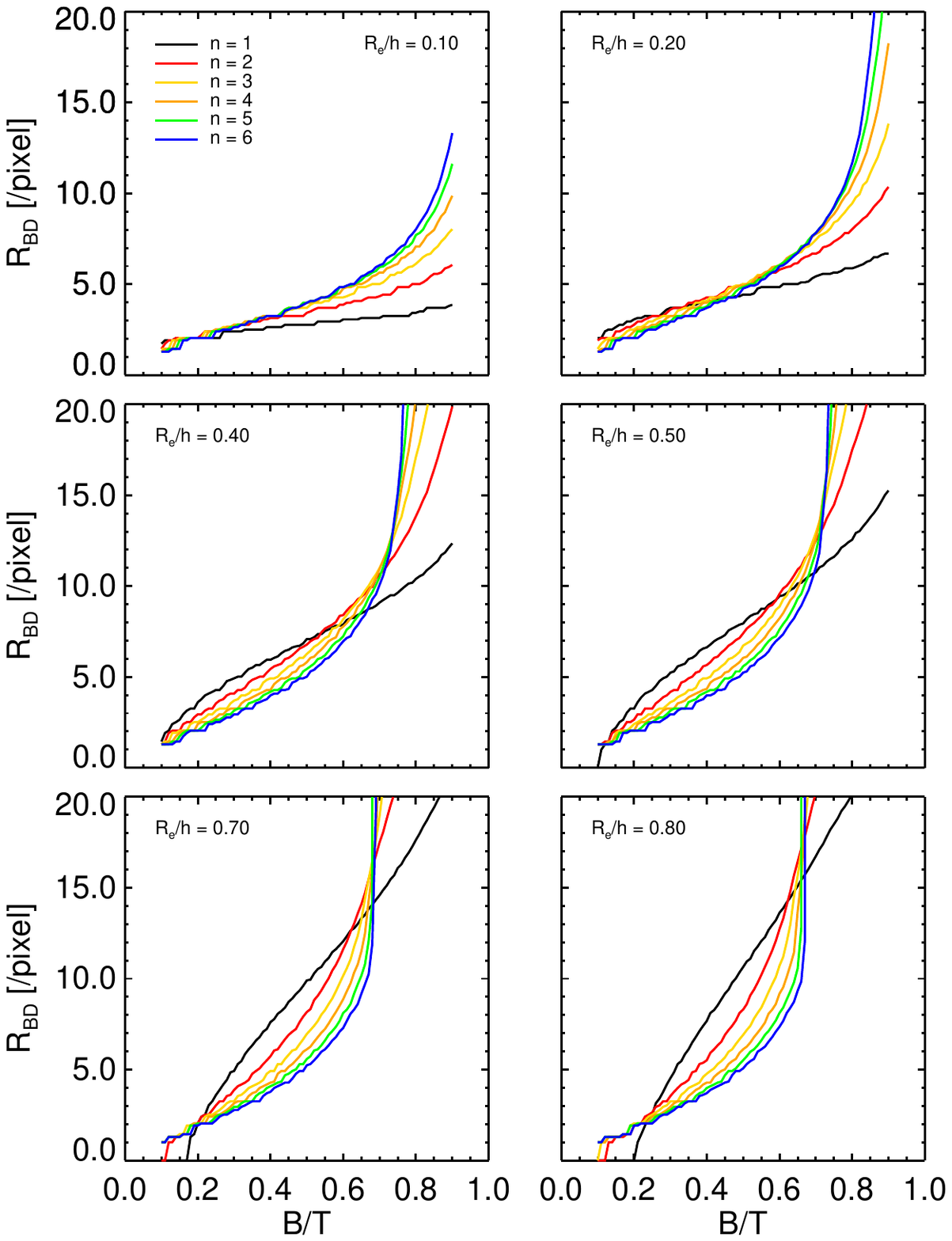}
\caption{Bulge extent $R_{BD}$ (see text) as a function of bulge-to-total luminosity ratio ($B/T$). Different colour lines represent different S\'{e}rsic indices and each panel shows a different $R_{e}/h$ ratio. For these synthetic galaxies $h=$10 pixels and $I_{0} = 10^{2.8}$ counts.}
\label{fig:R_BD}
\end{figure*}

We define the extent of the bulge region, $R_{BD}$, by subtracting the modelled disc-component from the modelled bulge-component and fitting an ellipse to the positive central pixels. This provides a good description of where the bulge is dominant in the bulge+disc galaxy model. Figure ~\ref{fig:R_BD} shows $R_{BD}$ as a function of $B/T$ for models with a range of values of $n$ (coloured lines) and $R_{e}/h$ ratio (different panels). 

The figure clearly illustrates the highly non-linear relation between the structural parameters and the extent of the bulge region. We see that galaxies with $B/T > 0.7$, $n > 2$ and $R_{e}/h>0.2$  will have the greatest number of data points available to fit. Understandably, bulges with $n=1$ will be more difficult to fit, as will bulges with low  $R_{e}/h$ or $B/T$. Interestingly, galaxies with high $B/T$ values will still have small bulge extents where $R_{e}/h$ is low. 

\begin{figure*}
\centering 
\includegraphics[width=1 \textwidth, trim=150 100 150 0, clip]{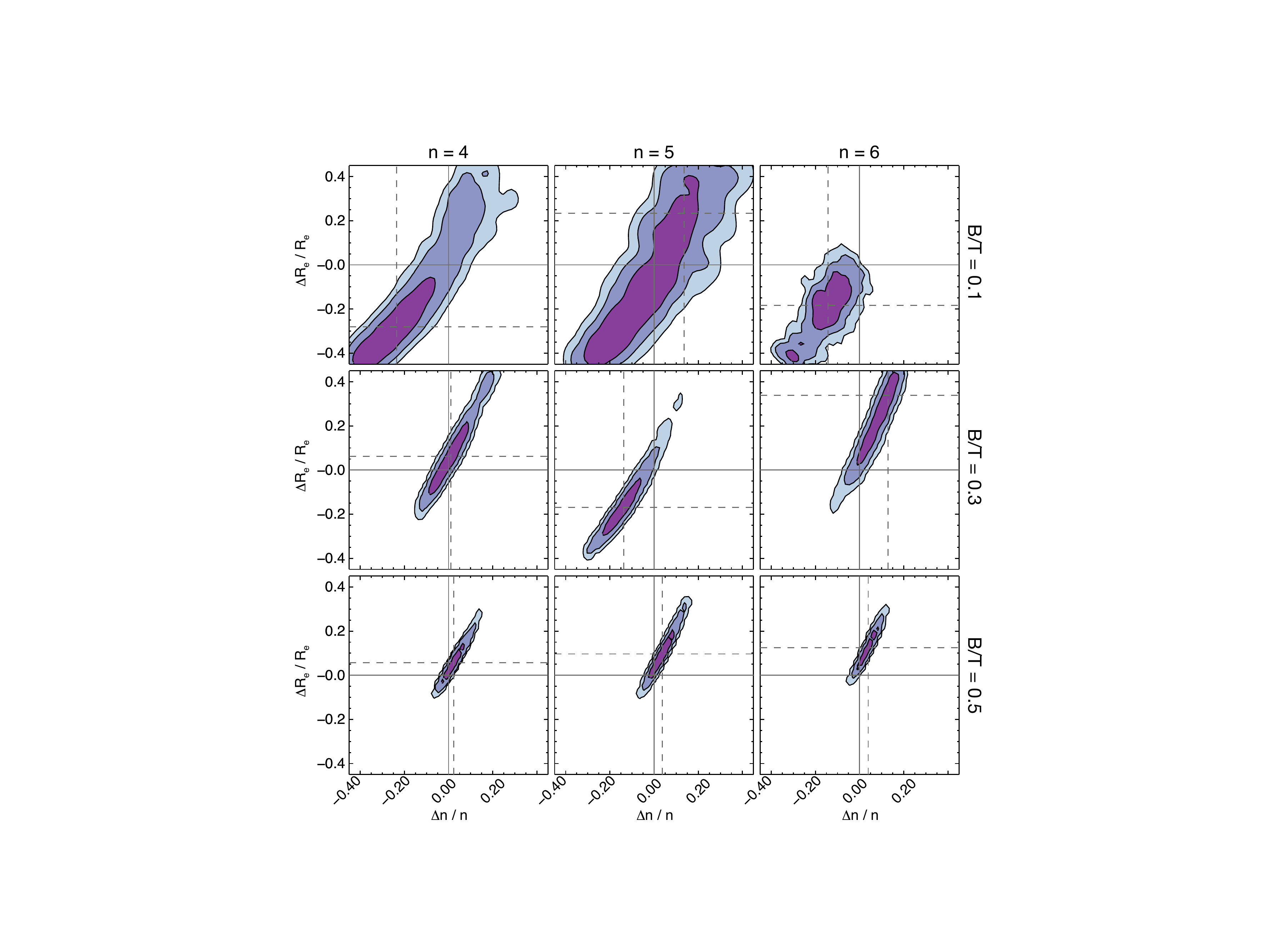}
\caption{The posterior fractional error distribution between $R_{e}$ and $n$ for nine synthetic galaxies. Each column (left to right) represents synthetic galaxies with true $n$ values of $n = 4, 5,$ and $6$, while each row (top to bottom) has a different $B/T$, $B/T = 0.1, 0.3, 0.5$. The contours are the $68\%$, $95\%$, and $99\%$ confidence regions. The dashed line shows the median of the distribution.}
\label{fig:re_vs_n_BT}
\end{figure*}

In Figure \ref{fig:re_vs_n_BT} we focus on galaxies with bulges with high-$n$ but low $B/T$ i.e. small bulges with profiles highly distinct from a disc-like structure. We show the posterior fractional error distribution in $n$ vs. $R_e$ for 9 synthetic galaxies with different values of $n$ and $B/T$. The figure shows the strong degeneracy between the fitted $n$ and $R_e$, resulting in very large errors on the median values, when $B/T$ is very small. This degeneracy can lead to biases in population fits, as found in Section \ref{sec:BulgeBias} and Fig.~\ref{fig:offn}. 

This is relevant, because it is commonly believed that galaxies with a dominant bulge (high $B/T$) component are more likely to have a higher $n$, while lower $B/T$ galaxies tend to have low $n$ values (e.g., this claim is demonstrated by Figure 11 of Gadotti 2009, and also by Figure 6 of Laurikainen et al. 2010, who use more complex decompositions of higher-resolution data). \nocite{Gad09, Lau10}. One implication of Fig. ~\ref{fig:R_BD} and Fig. ~\ref{fig:re_vs_n_BT} is that both these parameter combinations are easier to model, compared to galaxies with low $B/T$ and high $n$, or high $B/T$ and low $n$. This is a clear example of where studies need to be aware of degeneracies between parameters, and the potential for systematic biases. Future studies should be aware that the $B/T-n$ relation may be influenced by the resolution effects described here.

\normalsize
\newpage
\label{lastpage}

\end{document}